\documentclass[12pt,preprint]{aastex}
\def\dex#1{\hbox{$10^{#1}$}}
\def\tdex#1{\hbox{$\times10^{#1}$}}
\def\deg{$^\circ$}
\def\BMV{$B$$-$$V$}
\def\cmm#1{\hbox{${\rm cm^{-#1}}$}}
\def\fu{erg\,\cmm2\,s$^{-1}$\,\AA$^{-1}$}
\def\kms{km\,s$^{-1}$}
\def\mA{m\AA}
\def\Msun{M$_\odot$}
\def\Msunpyr{M$_\odot$\,yr$^{-1}$}
\def\HI{\protect\ion{H}{1}}
\def\Ha{H$\alpha$}
\def\He{H$\epsilon$}
\def\CII{\protect\ion{C}{2}}
\def\OI{\protect\ion{O}{1}}
\def\NaI{\protect\ion{Na}{1}}
\def\MgII{\protect\ion{Mg}{2}}
\def\SII{\protect\ion{S}{2}}
\def\CaI{\protect\ion{Ca}{1}}
\def\CaII{\protect\ion{Ca}{2}}
\def\FeI{\protect\ion{Fe}{1}}
\def\FeII{\protect\ion{Fe}{2}}
\def\vlsr{$v_{\rm LSR}$}
\def\vdev{$v_{\rm DEV}$}

\def\cahratio{\CaII/\HI\ ratio}
\def\cahratios{\CaII/\HI\ ratios}

\def\Smethod{2}
\def\Sobs{3}
\def\Smeas{4}
\def\Sresult{5}
\def\Sdisc{6}
\def\Sconcl{7}
\def\Fallsky{1}
\def\FCS{2}
\def\FPA{3}
\def\FPAdev{4}
\def\FGP{5}
\def\Fgp{6}
\def\FsCS{7}
\def\FsGP{8}
\def\Fsgp{9}

\begin{document}
\title{Distances to Galactic high-velocity clouds. I. Cohen Stream, complex~GCP, cloud~g1
\footnote{Based on observations carried out at the European Southern Observatory
(ESO), La Silla, under prog.\,ID No.\, 077.B-0359 (PI: P. Richter) with the UVES
spectrograph at the ESO Very Large Telescope, Paranal, Chile.}}

\author{%
B.P.\ Wakker\altaffilmark{1},
D.G.\ York\altaffilmark{2},
R.\   Wilhelm\altaffilmark{3},
J.C.\ Barentine\altaffilmark{4},
P.\   Richter\altaffilmark{5},
T.C.\ Beers\altaffilmark{6},
\v{Z}.\ Ivezi\'c\altaffilmark{7},
J.C.\ Howk\altaffilmark{8}}

\altaffiltext{1}{Department of Astronomy, University of Wisconsin, Madison, WI 53706; wakker@astro.wisc.edu}
\altaffiltext{2}{Astronomy \& Astrophysics Center, University of Chicago, Chicago, IL 60637; don@oddjob.uchicago.edu}
\altaffiltext{3}{Dept. of Physics \& Astronomy, Texas Tech University, Lubbock TX 79409; ron.wilhelm@ttu.edu}
\altaffiltext{4}{Dept. of Astronomy, Univ. of Texas, Austin TX 78712 and Apache Point Observatory, Sunspot, NM 88349; jcb@astro.as.utexas.edu}
\altaffiltext{5}{Institut f\"ur Physik, Universit\"at Potsdam, Am Neuen Palais  10, 14469 Potsdam, Germany; prichter@astro.physik.uni-potsdam.de}
\altaffiltext{6}{Department of Physics and Astronomy, CSCE: Center for the Study of Cosmic Evolution and JINA: Joint Institute for Nuclear Astrophysics, Michiga n State University, E.\ Lansing, MI 48824, USA; beers@pa.msu.edu}
\altaffiltext{7}{Dept. of Astronomy, University of Washington, Box 351580, Seattle, WA 98195; ivezic@astro.washington.edu}
\altaffiltext{8}{Dept. of Physics, University of Notre Dame, Notre Dame, IN 46556; jhowk@nd.edu}

%%%%%%%%%%%%%%%%%%%%%%%%%%%%%%%%%%%%%%%%%%%%%%%%%%%%%%%%%%%%%%%%%%%%%%%%%%%%%%%%

\begin{abstract}
The high- and intermediate-velocity interstellar clouds (HVCs/IVCs) are tracers
of energetic processes in and around the Milky Way. Clouds with near-solar
metallicity about one kpc above the disk trace the circulation of material
between disk and halo (the Galactic Fountain). The Magellanic Stream consists of
gas tidally extracted from the SMC, tracing the dark matter potential of the
Milky Way. Several other HVCs have low-metallicity and appear to trace the
continuing accretion of infalling intergalactic gas. These assertions are
supported by the metallicities (0.1 to 1 solar) measured for about ten clouds in
the past decade. Direct measurements of distances to HVCs have remained elusive,
however. In this paper we present four new distance brackets, using VLT
observations of interstellar \CaII~H and K absorption toward distant Galactic
halo stars. We derive distance brackets of 5.0 to 11.7~kpc for the Cohen Stream
(likely to be an infalling low-metallicity cloud), 9.8 to 15.1~kpc for
complex~GCP (also known as the Smith Cloud or HVC\,40$-$15+100 and with still
unknown origin), 1.0 to 2.7~kpc for an IVC that appears associated with the
return flow of the Fountain in the Perseus Arm, and 1.8 to 3.8~kpc for cloud~g1,
which appears to be in the outflow phase of the Fountain. Our measurements
further demonstrate that the Milky Way is accreting substantial amounts of
gaseous material, which influences the Galaxy's current and future dynamical and
chemical evolution.
\end{abstract}


%%%%%%%%%%%%%%%%%%%%%%%%%%%%%%%%%%%%%%%%%%%%%%%%%%%%%%%%%%%%%%%%%%%%%%%%%%%%%%%%

\section{Introduction}
\par The evolution of galaxies is strongly driven by the gas in the interstellar
medium. High- and low-mass stars form from gas in the disk and put back energy
into the gas. Some gas escapes from the disk, forming a hot atmosphere. As this
gas cools and rains back it forms a ``Galactic Fountain''. The hot atmosphere
has been called the ``Galactic Corona'', and it extends up to 10~kpc above the
Galactic disk. There is also infall of new material that provides fuel for
galaxy growth. This material can originate in small satellite galaxies, as gas
tidally pulled out of passing galaxies, and from more pristine intergalactic
gas. The cool clouds that fall in would be embedded in (or condense out of) a
much more extended (100--200~kpc radius) hot Corona, different from the one
formed by star-formation processes. Indirect evidence for infalling gas is
provided by two arguments (Chiappini et al.\ 2001). First, at the current rate
of star formation, all of the ISM would be turned into gas within about a Gyr.
Second, the narrowness of the distribution of metallicities of long-lived stars
is best explained if the metallicity of the ISM remains more or less constant
over a Hubble time, which happens if there is an inflow of low-metallicity
material from outside. The predicted inflow rate decreases with time and at
present is on the order of 1~\Msunpyr.
\par Direct observational evidence for the processes mentioned above is provided
by the high-velocity clouds (see reviews by Wakker \& van Woerden 1997; Richter
2006). They are defined observationally as clouds with velocities that are
incompatible with differential galactic rotation. In practice this means
$|$\vlsr$|$$>$50~\kms. Historically, such gas has been divided into the
high-velocity clouds (HVCs), having $|$\vlsr$|$$>$90 or $>$100~\kms, and the
intermediate-velocity clouds (IVCs), with $|$\vlsr$|$=50 to 100~\kms. They have
been mapped extensively in 21-cm emission (see Hulsbosch \& Wakker 1988; Wakker
\& van Woerden 1991; Hartmann \& Burton 1997; Morras et al.\ 2000; Kalberla et
al.\ 2005). The different origins of different clouds can be determined by
measuring their metallicities and distances. Solar metallicity clouds near the
disk trace the Galactic Fountain. Distant ($>$5~kpc), low-metallicity (0.01-0.2
solar) clouds trace infall. Long streams with intermediate metallicity are
likely to be tidal. Intrinsic metallicities ($Z$) have been determined for a few
HVCs (see van Woerden \& Wakker 2004 for a summary). The best values are
$Z$$\sim$0.15 times solar for complex~C (Wakker et al.\ 1999; Richter et al.\
2001; Tripp et al.\ 2003; Sembach et al.\ 2004), while clouds in the Magellanic
Stream have SMC-like metallicity ($\sim0.25$ solar; Lu et al.\ 1998; Gibson et
al.\ 2000). Van Woerden \& Wakker (2004) list approximate values for a total of
eight HVCs, ranging from 0.1 to 1 times solar. In contrast, the
intermediate-velocity clouds are usually found to have near-solar metallicity
(Wakker 2001; Richter et al.\ 2001).
\par The main unknown for the HVCs remains their distances. If we knew these, we
could derive empirical (rather than theoretical) estimates for the rate of
accretion of low-metallicity (0.01--0.1 solar) gas, the rate of circulation of
gas between disk and halo,, and the mass of tidal material around the Milky Way.
Furthermore, HVCs could be used to calibrate the fraction of ionizing photons
escaping the Milky Way disk, as many HVCs are seen in \Ha\ emission (see
Bland-Hawthorn \& Maloney 1999).
\par Although the first paper on determining distances to HVCs was published
four decades ago (Prata \& Wallerstein 1967), progress has been slow. Wakker
(2001) summarized the state of knowledge a few years back. It is clear that
distances to several IVCs are on the order of 1--2~kpc. However, for HVCs Wakker
(2001) only lists lower limits of 6~kpc to complex~C (\vlsr$\sim$$-$150~\kms),
5~kpc to complex~H (\vlsr$\sim$$-$180~\kms), and 0.3~kpc to the Anti-Center
clouds (clouds with \vlsr$\sim$$-$250~\kms\ and \vlsr$\sim$$-$100~\kms), as well
as two upper limits -- 10~kpc for the high-latitude part of complex~A
(\vlsr$\sim$$-$150~\kms), based on the RR\,Lyrae star AD\,UMa (van Woerden et
al.\ 1999), and 13~kpc for complex~WE (\vlsr$\sim$+110~\kms), using the O9.7II
star HD\,156359 (Sembach et al.\ 1993). Since then, Wakker et al.\ (2003) has
reported a lower limit of 8~kpc for complex~A, while Thom et al.\ (2006) used
the star HE\,1048+0231 to set an upper limit of 8.8~kpc to cloud WW\,35
(\vlsr$\sim$+100~\kms).
\par In this paper we report on our observations of 25 targets in the direction
to two HVCs and two IVCs. These yield distance brackets to all four clouds. We
describe the method used to derive the distances of these clouds in
Sect.~\Smethod. Sect.~\Sobs\ summarizes our observations, while the manner in
which we make our measurements is discussed in Sect.~\Smeas. Section~\Sresult\
presents the results for each of the four clouds, describing the distance
determination, notes on each of the stars, a summary of \cahratios, a discussion
of the implied location in the Milky Way and a discussion of their ionized
hydrogen content. In Sect.~\Sdisc\ we present a short general discussion of our
results, while in Sect.~\Sconcl\ we summarize the conclusions.

%%%%%%%%%%%%%%%%%%%%%%%%%%%%%%%%%%%%%%%%%%%%%%%%%%%%%%%%%%%%%%%%%%%%%%%%%%%%%%%%

\section{Determing HVC/IVC distances}
\subsection{Basic method}
\par The distance to an HVC/IVC can be determined by looking for interstellar
absorption at the velocity of the cloud in spectra of stars with known
distances. A detection sets an upper limit, while a {\it significant}
non-detection sets a lower limit. A significant non-detection means that the
ratio of the expected equivalent width to the observed upper limit is
sufficiently large (e.g.\ $>$10; the exact value depends on the ion that is
observed, and on knowledge of the cloud's ionic abundance, see Wakker 2001). The
best kind of stars for this purpose are Blue Horizontal Branch (BHB) and
RR\,Lyrae stars (in their hot phase). This is because they are fairly numerous,
their distances can be determined relatively well, and they are relatively blue,
resulting in few confusing stellar lines. This method is described in detail by
Wakker \& van Woerden (1997) and Wakker (2001).
\par There are three steps in the method: (1) locate candidate probe stars, (2)
obtain intermediate-resolution spectra and photometry to derive their distances,
velocities and abundances, and (3) obtain high spectral resolution (15~\kms\ or
better) and high signal-to-noise ratio spectra for stars at a range of
distances; these stars should be selected to have minimal blending between
stellar and interstellar lines. In this paper, we only summarize steps (1) and
(2), while we report on the most extensive set of step (3) observations
presented in the literature so far. Our extensive program covering steps (1) and
(2) will be described in subsequent papers.

\subsection{Stellar probes}
\par Until recently, most of the candidate stellar probes were found from three
catalogues: the variable star lists of Kukarkin (1969, 1970), the PG survey
(Green et al.\ 1986) and the HK survey (Beers et al.\ 1996). These list many BHB
and RR\,Lyrae candidates with distances up to about 5~kpc, as well as a few more
distant stars (up to $\sim$10~kpc). The recent data from the Sloan Digital Sky
Survey (SDSS) and the 2-Micron All-Sky Survey (2MASS; Cutri et al.\ 2003) now
allow us to make further progress, adding candidate stars at up to 100~kpc
distance.
\par The SDSS project and the creation of the archive are discussed by York et
al.\ (2000). The details of the data products are described by Stoughton et al.\
(2001) and by Adelman-McCarthy et al.\ (2005 and references therein). A
2.5-meter telescope (Gunn et al.\ 2006) is equipped with two dual-beam
spectrographs (R$\sim$1800), operating between 3900~\AA\ and 9000~\AA, and with
a CCD camera (Gunn et al.\ 1998) with a 5 filter photometric system (Fukugita et
al.\ 1996), calibrated on imaging nights with a 20-inch Photometric Telescope
(Hogg et al.\ 2001, Smith et al.\ 2002; Ivezic et al.\ 2004; Tucker et al.
2006). The CCD camera is used to scan the northern sky in the five filters, from
which our RR\,Lyrae or BHB candidate stars are chosen, using color criteria
(Sirko et al.\ 2004; Ivezi\'c et al.\ 2005). The star positions are determined
in the manner described by Pier et al.\ (2003). These stars have distances
ranging from $\sim$5~kpc up to 100~kpc, as estimated from their $V$ magnitudes
($V$=0.56$r$+0.44$g$+0.056; Ivez\'ic et al.\ 2000) and assuming an absolute
magnitude of $M_V$=0.7 for RR\,Lyraes (Layden et al.\ 1996) and $M_V$=1.040 +
(\BMV)*($-$4.423+(\BMV)*(17.74+(\BMV)*($-$35.73))) for BHBs (Preston et al.\
1991).
\par The 2MASS survey scanned the sky at 2\arcsec\ pixels, using automated 1.3m
telescopes on Mt.\ Hopkins and CTIO, taking images at $J$, $H$, and $K$. It
includes thousands of BHB candidates with distances up to $\sim$8~kpc, which can
be selected from their $J$$-$$H$ and $J$$-$$K$ colors (Brown et al.\ 2004).
\par For some stars the SDSS provides intermediate-resolution ($\sim$2~\AA)
spectra. To obtain similar spectra for more SDSS stars, as well as for the 2MASS
selected stars, we have a large observing project underway at the 3.5-meter
telescope at Apache Point Observatory (APO). Between 2003 and 2006 we obtained
spectra at APO for some 200 stars.
\par Using these spectra, we can derive the effective temperature and gravity
for each star, following the method outlined by Wilhelm et al.\ (1999).
Combining with photometry these parameters can be used to derive good distance
estimates. We will describe these observations (as well as improvements to the
original method) in a separate paper (Wilhelm et al.\ 2007, in preparation).
Here we just present an outline.
\par Our method compares the observed photometry (either SDSS $ugr$ or Johnson
$UBV$), and observed spectroscopy. We compare the Balmer line widths and
equivalent width of the \CaII~K line with a grid of sythetic colors and
spectroscopy parameters that are generated from {\it Atlas9} model atmospheres.
Combining the Balmer line widths with $u$$-$$g$ (or $U$$-$$B$) color indices
allows the determination of $\log g$ values across the temperature range
6000$<$$T_{\rm eff}$$<$12000 K. The determined stellar parameters ($T_{\rm
eff}$, $\log g$ and [Fe/H]) are then used to estimate absolute magnitudes $M_V$
by placing the star on the theoretical isochrones of Girardi et al.\ (2004).
This method allows the determination of distances for stars of various
metallicities on the zero-age horizontal branch, on the main-sequence, as well
as for terminal HB stars and post-main sequence blue stragglers. We estimate
errors in the stellar parameters by changing the input values for observables
(i.e.\ line widths and colors) by $\pm$1$\sigma$ one observable at a time. We
thus recalculate $T_{\rm eff}$, $\log g$, [Fe/H], $M_V$, and $D$ many times. The
error follows as the rms spread in the set of results for each parameter.

\subsection{Implementation}
\par The results of the programs described above yield a sample of stars with
known distances in the direction of HVCs. These then need to be observed with
high spectral resolution ($R$$>$20000), in order to separate the different
interstellar absorption components, as well as to maximize the sensitivity to
the $\sim$20~\kms\ wide HVC/IVC absorptions. In principle, \MgII
$\lambda\lambda$2796.352, 2803.531 and \CII $\lambda\lambda$1036.337, 1334.532
are the best lines, as they saturate at low \HI\ column densities
($\sim$7\tdex{17}\,\cmm2). However, they require sparsely available time on
telescopes in space, which are not sensitive enough to observe many 16th to 18th
magnitude BHB stars.
\par In the optical the strongest absorption lines are the \CaII
$\lambda\lambda$3934.770, 3969.591 K and H lines. Previous observations of
\CaII\ absorption in HVCs with bright 21-cm emission have yielded six
detections, giving \cahratios\ of 17$\pm$6\tdex{-9} for complex~A (Schwarz et
al.\ 1995), 22$\pm$1\tdex{-9} for complex~C (Wakker et al.\ 1996),
3.3$\pm$0.9\tdex{-9} for complex~M (Bowen et al.\ 2000), 33$\pm$4\tdex{-9} for
the Magellanic Stream (Songaila \& York\ 1981), 66$\pm$8\tdex{-9} for the
Leading Arm of the Magellanic Stream (West et al.\ 1985) and
270$\pm$120\tdex{-9} for complex~WB (Robertson et al.\ 1991). There are also
several \CaII\ detections toward extragalactic supernovae in directions with
very faint or no 21-cm (d'Odorico et al.\ 1985, 1989; Meyer \& Roth 1991;
Vladilo et al.\ 1994; Ho \& Filippenko 1995, 1996), which yield very high
\cahratios. Toward IVCs \cahratios\ range from 8\tdex{-9} to 55\tdex{-9} in the
IV Arch, from 12\tdex{-9} to 17\tdex{-9} in the LLIV Arch, 90\tdex{-9} in
complex~K, 50\tdex{-9} in the PP Arch, and 65\tdex{-9} in complex~gp (Wakker
2001).
\par Using these detections of high- and intermediate-velocity absorption (as
well as HVC/IVC absorption in other ions, such as \FeII) Wakker \& Mathis (2000)
discovered that for \CaII, \FeII, etc.\ the [ion]/\HI\ ratio correlates with
$N$(\HI). The physical origin of this effect remains unexplained, as discussed
by Wakker \& Mathis (2000). For \CaII\ this correlation can be expressed as:
$$ R({\rm CaII})\ =\ N({\rm CaII})/N({\rm HI})\ =\ -0.78\ (\ \log(\ N({\rm HI})\ -\ 19.5\ )\ )\ -\ 7.76. $$
In fact, this relation is a better predictor of the \cahratio\ than the
assumption that $R$(\CaII) is constant in a particular cloud. Therefore, we use
this relation to convert the known \HI\ column densities in HVCs and IVCs in the
direction of a candidate target star to predict the \CaII\ equivalent width.
This allows us to construct a sample of stars for each HVC by selecting the ones
with the largest predicted equivalent widths in a set of distance intervals
(e.g.\ 1--3~kpc, 3--6~kpc, 6--9~kpc, etc.). We have done this for sixty HVC/IVC
fields. In this paper we report on observation of stars in three fields,
covering two HVCs and two IVCs.
\par Previous authors have set distance brackets or limits to HVCs/IVCs using
the nominal distances to stars that give upper limits and significant lower
limits. However, the errors in the stellar distances have been ignored in the
past. In this paper we take them into account. Since the error distribution is
centered on the nominal distance, the probability that the upper limit is less
than the {\it nominal} distance to the star is 50\%. If the error distribution
of the stellar distances were gaussian, the probability that the upper limit is
less than $D_{\rm nom}$+0.47$\sigma(D)$ is 68\%. Thus, the 68\% (i.e. 1$\sigma$)
confidence interval on the upper distance limit to the HVC is given by $D_{\rm
nom}$+0.47$\sigma(D)$. A similar reasoning provides the lower limit of the
bracket. At present, we do not calculate the actual error distribution for the
distance, so for convenience we assume it is gaussian. For future work we will
improve this step.

%%%%%%%%%%%%%%%%%%%%%%%%%%%%%%%%%%%%%%%%%%%%%%%%%%%%%%%%%%%%%%%%%%%%%%%%%%%%%%%%

\section{Observations}
\subsection{Stellar data}
\par For the April-September 2006 semester we were given an allocation of 37
hours on ESO's Very Large Telescope (VLT) to observe 24 stars and one QSO
(proposal ID 077.B-0359). For our observations we used the UVES spectrograph
installed on the VLT together with a 1\arcsec\ slit and the blue arm (CD2). This
setup provides a wavelength coverage from $\sim$3730 to 4990~\AA, a pixel size
of 0.03~\AA\ ($\sim$2.3~\kms), and a spectral resolution of $R$$\sim$40,000,
corresponding to an FWHM of $\sim$7.5~\kms. During our observations the seeing
typically was $\leq$0\farcs8. The raw data were reduced with the standard UVES
pipeline, which includes flat-fielding, bias- and sky-subtraction, and a
relative wavelength calibration. To compare with the 21-cm data, the final
spectra were then shifted to an LSR velocity scale. For multiple integrations of
the same background source, the individual spectra were co-added, weighted by
their inverse variances. In the continuum the signal-to-noise ratio (S/N) per
resolution element in the UVES data typically ranges from about 40 to 100.
\par The three HVCs that we selected for our VLT observations have in common
that they are above the horizon during the southern winter. They are the ``Cohen
Stream'', ``complex~GCP'' and ``cloud~g1''. Most stars in the Cohen Stream field
also lie projected onto another, more nearby IVC. We describe these clouds and
the stars projected on them in the next section. Figure~\Fallsky\ shows the
locations of these three clouds on a map of the HVC sky. The map also shows the
three stars toward which HVC absorption was found previously.
\par Table~1 presents the stellar data. Coordinates are defined in the names for
2MASS and SDSS stars, given in the notes for the remaining stars. The two ``BS''
stars are from the HK survey (Beers et al.\ 1996) and the numbers represent
plate number and star id on that plate. Seventeen stars in the VLT sample have
classification spectra from our APO observations, one spectrum is from the SDSS
(see Col.~2 of the table). We also obtained photometry for most stars, with the
source of the data listed in Col.~3 and the resulting magnitudes in Cols.~4, 5
and 6. The extinction (from Schlegel et al.\ 1998) is given in Col.~7. For five
stars we can use SDSS photometry. For twelve stars (eleven from 2MASS, one from
the HK survey), we obtained this photometry ourselves. For seven stars we do not
have photometry data. Three of these are 2MASS stars, so we used the star's B
magnitude from the 2MASS catalogue. For the three named RR\,Lyrae stars we used
the photometry listed in the Kukarkin et al.\ (1970) catalogue. Finally,
Str\"omgren photometry was obtained by Moehler et al.\ (1990) for PG\,0142+148.
\par For all stars we give the stellar parameters derived directly from the
high-resolution spectra, while for the 17 stars with intermediate-resolution
spectroscopy we give additional lines showing the stellar parameters resulting
from our APO data, for comparison. The stellar parameters are derived using the
photometry, the Balmer lines, \CaII~K and H and \CaI$\lambda$4227.918. A full
description of this method will be given by Wilhelm et al.\ (2007, in
preparation). Columns~8 to 11 list the resulting effective temperature, gravity,
metallicity and stellar type, respectively. Combining the stellar parameters
with photometry and isochrones, we can derive the star's distance, given in
Col.~13 of the table.
\par The VLT data were used to derive the stellar velocities listed in Col.~12.
This was done by fitting a gaussian to the stellar \CaII, \CaI\ and \FeI\ lines,
then averaging the fitted central velocities. For some of the stars the stellar
velocity varies between exposures (see the notes to the Table). In these cases,
we did not use all of the exposures, but only the ones with the least confusing
separation between stellar and interstellar lines. The listed velocity is valid
for that spectrum.
\par There are often differences in the temperature, gravity and metallicity
derived from the VLT spectrum when compared to those values derived from the APO
spectra. When the differences are more than one sigma, they are noted in the
following subsection.

\subsection{Distance estimation}
\subsubsection{Non-problematic stars}
\par For eight stars (2MASS\,J014936.48+143914.6, 2MASS\,J021409.31+090105.3,\\
2MASS\,J195927.29+000822.3, 2MASS\,J195823.36-002719.0,
2MASS\,J195912.00-002645.3,\\ 2MASS\,J213327.04+133026.7,
2MASS\,J213520.04+133045.0, BS\,17578-0016), the determination of the stellar
distance based on the VLT and APO spectra is the same to within 15\%. For
four stars (2MASS\,J014825.88+132305.3, SDSS\,J014843.61+130411.5,
2MASS\,J195922.75+000300.0 and 2MASS\,J213451.58+134017.5 the determinations
differ by 20--40\%. Since they are still the same within $\sim$1$\sigma$, the
results are still compatible, just relatively inaccurate.
\par In the case of 2MASS\,J213451.58+134017.5 the metallicity indicated by the
\CaII~K line is about 1 dex less than that found from spectral regions with many
metallic lines (mostly Fe). The star is probably a main-sequence star with solar
abundance, so we adopt a nominal $\log g$=4.0 to derive the distance. The lower
resolution of the APO data makes that result less reliable, so we are not
worried by the $\sim$1$\sigma$ difference in the nominal distance.
\par For one star (2MASS\,J195925.49$-$000519.1) the difference between the
distance derived from the high- and the intermediate-resolution spectrum is
larger than the nominal error (1.5$\pm$0.2~kpc vs 2.6$\pm$0.4). However, these
values are still somewhat consistent, and this is not an important star for our
program to determine HVC distances. We therefore did not try to correct his
discrepancy.
\par For two stars (SDSS\,J015735.65+135254.2, and BS\,17578-0015), we have no
APO classification spectrum, but we do have photometry, and we are confident
that the resulting classifications and distances are good.
\par Two stars are too hot (i.e.\ $T$$>$12000~K) for our method, but distances
are given in the literature. These are PG\,0142+148, which was analyzed by
Moehler et al.\ (1990) and PG\,2134+125, which is the central star of the
planetary nebula NGC\,7094. Phillips (2004) summarizes distance determinations
for planetaries, and includes NGC\,7094.

\subsubsection{Problematic stars (non-RR\,Lyrae)}
\par For two stars the distances derived from the VLT and APO spectra differ
substantially.
\par The VLT spectrum of 2MASS\,J021651.34+080150.2 shows many stellar lines,
which blend in the APO spectrum, making it difficult to determine the gravity.
In addition we do not have photometry for this star, so that we cannot directly
fit $T_{\rm eff}$ and $\log g$ (see Wilhelm et al.\ 2007). We instead use a grid
of spectral models with different temperatures and gravities, and find the best
match. As a result of these problems, the APO spectrum suggests low gravity,
while the VLT spectrum indicates high gravity. This leads to a dichotomy in the
classification; we have more confidence in the lower (VLT-based) distance.
\par The VLT spectrum of 2MASS\,J195741.61-004009.7 also shows many stellar
lines. In fact, the line density is such that it explains why the APO spectrum
of this star is noisy, making its analysis difficult. The determination of the
star's gravity is so uncertain that the APO data suggest it is a BHB, implying
$D$11.1$\pm$3.3~kpc, while the VLT spectrum indicates a main-sequence F star at
0.7$\pm$0.6~kpc. We judge that it is more likely to be nearby than to be
distant.

\subsubsection{RR\,Lyrae stars}
\par The final five stars are RR\,Lyrae stars, a classification based on the
derived stellar parameters, but also on the fact that we can see the stellar
lines shift between different VLT exposures. To estimate a distance for an
RR\,Lyrae star requires a determination of its average magnitude. We usually do
not have phase information, so we do not know whether the VLT spectrum or the
SDSS photometry was taken during maximum, minimum or somewhere between.
Moreover, the magnitudes given by Kukarkin et al.\ (1970) for V1084\,Aql,
V1172\,Aql, and FW\,Peg are photographic, not $UBV$. They are also quantized, in
steps of $\sim$0.25 magnitudes. Therefore, we use the following reasoning to
estimate the distances to the five RR\,Lyrae stars in our sample.
\par We start by looking at SDSS\,J015133.91+141105.2. The SDSS photometry of
this star gives $V$=16.81 (using the transformation from $ugriz$ to $UBV$ given
by Ivezi\'c et al.\ 2000). We have six VLT exposures of this star, taken 28 and
16 days apart, with one run of four exposures in four consecutive hours. The
star's flux varies between 6.2\tdex{-16} \fu\ and 20.5\tdex{-16} \fu, giving
13.3\tdex{-16} \fu\ on average, corresponding to $V$=16.14. Since the range in
flux is about 1.3 magnitude, it is reasonable to conclude that 16.1$\pm$0.2 is a
good estimate for the average magnitude of this star. Combined with an absolute
magnitude for RR\,Lyraes of $M_V$=0.7 (Layden et al.\ 1996) and extinction
$A_V$=0.16 (Schlegel et al.\ 1998), this implies a distance of 11.2$\pm$1.0~kpc.
\par The flux of SDSS\,J020033.49+141153.9 varies from 10.7\tdex{-16} to
21.6\tdex{-16} \fu\ between the three spectra of this star. The average flux is
about 16\tdex{-16} \fu, or $V$$\sim$16.0. The temperature that is derived is
about 6600~K, which suggests the star is in its cool phase. The SDSS equivalent
$V$ magnitude is 15.58. These results are compatible if the SDSS observation
caught it near average while the VLT observation caught it near minimum. Based
on this, we set the average magnitude to $V$=15.6$\pm$0.3. With $A_V$=0.19 this
implies $D$=8.7$\pm$1.3~kpc.
\par For V1084\,Aql Kukarkin et al.\ (1970) gave $<$$P$$>$=17.0. Bond (1978)
included this star in a sample of RR\,Lyraes for which he obtained photometry.
He finds $<$$V$$>$=17.0, with an accuracy of about 0.2 mag. The star's flux
varies from 7.4\tdex{-16} to 14.5\tdex{-16} \fu\ in our four VLT exposures. On
average this flux corresponds to $V$=16.35. We derive a temperature of 7000~K,
which suggests we observed the star near maximum phase, which is compatible with
the relatively high flux and the relatively low number of stellar lines. Using
$A_V$=0.49 (Schlegel et al.\ 1998) then implies a distance of 14.5$\pm$1.3~kpc.
\par Kukarkin et al.\ (1970) give an average magnitude for V1172\,Aql of
16.5$\pm$0.25. We find an average flux in the VLT spectra that corresponds to
$V$=16.36. The temperature that is derived from the three VLT exposures is about
6500~K, suggesting we observed it in a cooler, fainter phase. We thus assume
that the Kukarkin et al.\ (1970) magnitude is slightly low and calculate the
distance using $<$$V$$>$=16.4$\pm$0.3, which for $A_V$=0.60 gives
$D$=10.5$\pm$1.4~kpc.
\par The star FW\,Peg has $<$$V$$>$=14.0 in Kukarkin et al.\ (1970). The VLT
spectrum gives a flux corresponding to $V$=14.68, and a temperature of 6500~K.
This is compatible if we caught the star near minimum. We thus use
$<$$V$$>$=14.0$\pm$0.5 to calculate its distance. With $A_V$=0.40 this gives
3.8$\pm$0.8~kpc.

\section{Measurements}
\par We now summarize the measurements we made on the stars, and then discuss
for each cloud in turn our previous knowledge, followed by a more general
discussion of our results concerning their distances, \cahratios\ and location
in the Milky Way.

\subsection{Stellar measurements}
\par Table~2 presents the stellar data that pertain to searching for the
interstellar lines. We give the star's Galactic location in Cols.~2 and 3, and
repeat the distance from Table~1. Column~5 gives the total VLT exposure times
(but see the notes to the table for exposures that were excluded for the final
analysis). We also determined the star's continuum flux, as well as the flux and
S/N ratio at the velocity of the high-velocity clouds (Cols.~6, 7, 8). This was
done by fitting a polynomial to the line-free regions of the continuum. To
calculate the flux at the HVC's velocity, the stellar \CaII\ absorption was
considered as ``continuum'', since we are interested in the flux and S/N
appropriate for measuring the interstellar line against the background of the
stellar flux. The order of the fitted polynomial lies between 1 and 4. The
signal-to-noise ratio is calculated per 7~\kms\ resolution element, i.e.\
calculated after rebinning the spectrum to three pixels. It is the ratio of the
flux to the rms of the fit around the line-free regions of the spectrum used in
the fit. This value is compared with the nominal error in the flux to make sure
that it is similar.

\subsection{\HI\ and \CaII\ measurements}	
\par Table~3 lists the \CaII\ interstellar measurements. The table contains the
following columns.
\par Columns~1 and 2 repeat the stellar name and distance.
\par Columns~3 and 4 give the velocity and the logarithm of the \HI\ column
density of the HVC, IVC, or low-velocity gas in the sightline. The \HI\ 21-cm
emission data were taken from the Leiden-Argentina-Bonn Survey (LAB; Kalberla et
al.\ 2005). This survey provides spectra on a 0\fdg5$\times$0\fdg5 grid, with a
36\arcmin\ beam. We calculated a weighted average of all spectra that lie within
0\fdg6 from each star as the best approximation to the \HI\ spectrum toward each
star. To determine $v$(\HI) and $N$(\HI) we fitted gaussians to these spectra,
using one gaussian component for the high- or intermediate-velocity emission,
and between two and four for the low-velocity gas. It should be noted that the
intermediate- and high-velocity components usually are not exactly gaussian, so
the fit gives only an approximation to $N$(\HI).
\par The uncertainty in $N$(\HI) is dominated by the fact that the Dwingeloo
telescope has a large beam on the sky. Wakker et al.\ (2001) showed the value of
$N$(\HI) measured with a 9 arcmin beam can be up to a factor of 2.5 lower or
higher than the value measured with a 35 arcmin beam. Further, when comparing
measurements of $N$(\HI) made with a 9 arcmin beam with those made with a 1--2
arcmin  beam, the latter are between 75 to 125\% of the former. The origin of
these differences is clearly small-scale structure in the cloud, including edge
effects if the beam is near a cloud's edge. The distribution of the ratio
$N$(\HI,real)/$N$(\HI,LAB) has an average of about 1, and an rms of about 0.2.
As a result, we assume a 20\% error for $N$(\HI) when calculating \cahratios.
Ideally, we would like to obtain high-resolution (1 arcmin or better) \HI\ 21-cm
data toward each of our targets, but we do not have such data at this time. Even
then, there is always the possibility of a hole in the distribution of the \HI\
in the direction of the star, and a non-detection would still not imply that the
star lies in front of the HVC. However, we use to the apparent hierarchical
structure of the \HI\ in HVCs (see Schwarz \& Wakker 2004) to argue that it is
possible to account for the uncertainty in $N$(\HI) by using a ``safety
factor'', as described below under the discussion of Col.~13.
\par Columns~5 and 6 give the predicted K (Col.~5) and H (Col.~6) equivalent
widths. These are derived from a predicted \CaII\ column density, combined with
the observed linewidth of the \HI\ emission. The prediction for the column
density is derived from the correlation between $N$(\CaII) and $N$(\HI)
discovered by Wakker \& Mathis (2000), but with a cloud-dependent correction
factor. The correction factor is found by comparing the nominal prediction
(given a value of $N$(\HI)) for the \cahratio\ with the actual observed value,
and then averaging the resulting quotient between all detections found for each
cloud. We will discuss the correction factors in Sect.~\Smeas, but note here
that the quotient of nominal prediction and actual value tends to be a constant
value in a particular cloud, though it varies between clouds.
\par Columns~7 and 8 list the measured K and H line equivalent width detections
or 3-$\sigma$ limits. For detections the absorption lines were integrated over
the range of velocities where the flux lies below the continuum, which is
20--25~\kms\ wide for the Cohen Stream, the IVC projected onto the Cohen Stream,
and for complex~GCP, and about 40~\kms\ wide for cloud~g1. The equivalent width
limits for non-detections were determined by integrating the spectrum over a
20~\kms\ velocity range, and setting the upper limit to three times the one
sigma error in the equivalent width. The equivalent width error includes the
uncertainty associated with the noise, as well as the uncertainty associated
with continuum placement.
\par In Cols.~9 and 10 we give the \CaII\ column density derived by integrating
the apparent optical depth profile (Savage \& Sembach 1991): $$
N = \int N_a(v) dv = {m_e c\over \pi e^2}\,f\,\lambda\ \ln{C(v)\over F(v)},
$$ where $C(v)$ is the continuum and $F(v)$ the observed flux.
\par The resulting column densities are divided by $N$(\HI), giving the values
in Cols.~11 and 12. We note that the error in the \cahratio\ is calculated using
the observed error in the \CaII\ column density combined with an assumed 20\%
error in $N$(\HI).
\par Column~13 shows the final result derived from each star: an ``U'' if the
star yields an upper distance limit, an ``L'' if it gives a lower limit and a
blank if the S/N ratio was insufficient to derive a lower limit from a
non-detection. The significance of the non-detections is obtained by comparing
the expected value (Cols.~5 and 6) with the 3$\sigma$ upper limits (Cols.~7 and
8). As discussed by Wakker (2001), a safety factor needs to be taken into
account for such a comparison, since we do not precisely know the \HI\ column
density (see discussion of Col.~4 above) and the \cahratio\ in the cloud may in
principle vary due to variations in calcium depletion and ionization conditions.
We assume a factor $\sim$1.5 for the uncertainty in $N$(\HI) and another factor
$\sim$2 for the uncertainty in the Ca abundance. Only if the ratio
EW(expected)/$\sigma$(EW) is $>$9 do we conclude that a non-detection sets a
lower limit to the cloud's distance.

\subsection{Calculation of masses and mass flow rates}
\par After we determine the distance to a cloud, we can estimate its location in
the Milky Way, its mass, and the mass flow rate associated with it. These
numbers are summarized in Table~4.
\par We estimate the mass in the manner described by Wakker \& van Woerden
(1991). In summary, we find the values of T$_{\rm B}$ for each point in the
cloud that were given by Hulsbosch \& Wakker (1988). Converting T$_{\rm B}$ to a
column density (using the linewidth), and then multiplying by the area (in
\cmm2) represented by one survey grid cell (about one square degree) gives the
\HI\ mass in the grid cell. Integrating over the cloud gives the total mass,
given an assumed distance, $D$. This value scales as $D^2$.
\par We use the Hulsbosch \& Wakker (1988) dataset instead of the newer LAB data
from Kalberla et al.\ (2005), because in the latter we have to select a fixed
velocity range, so there can be blending with lower-velocity gas. Also, the
cloud edges need to be defined by polygons. The older survey clearly separates
any particular HVCs from other gas. An improved list could be constructed from
the LAB survey, but this has not yet been done. Where possible we do check the
Hulsbosch \& Wakker (1988) cloud masses against a rough estimate using the LAB
survey, and find that they are similar to about 25\%.
\par To derive the mass flow rate, we need to estimate a vertical velocity for
each survey grid cell. Given the observed LSR velocity, we calculate the
``deviation velocity'' (\vdev), the amount by which the cloud's velocity differs
from the maximum velocity that can be understood from a simple model of
differential galactic rotation (see Wakker 1991). This simple model consists of
a disk with radius 26~kpc with thickness 4~kpc inside the solar circle (8.5~kpc)
increasing to 12~kpc at 26~kpc. The rotation curve is flat, except in the inner
0.5~kpc where solid-body rotation is assumed. Using the deviation velocity we
estimate the cloud's $z$-velocity ($v_z$) by assuming that it is purely vertical
or by assuming that it is purely radial with respect to the Sun. In the former
case, $v_z$=\vdev/$\sin b$, while in the latter $v_z$=\vdev$\sin b$. To estimate
the mass flow, we take the average of these two possibilities. Dividing the
cloud's distance by the estimated vertical velocity yields the time it will take
for the cloud to reach the Galactic plane. Finally, the quotient of cloud mass
and travel time gives the mass flow associated with the cloud. The resulting
number scales as $D$.
% Mdot = M/t = mD^2 vz/D t = D/vz

\subsection{Ionized hydrogen}
\par To fully understand the clouds and estimate the mass flows associated with
them, we also need to estimate the amount of ionized hydrogen. It turns out that
\Ha\ data exists for each of our clouds. There are a number of single pointings
for complex~GCP and the Cohen Stream, and there are maps from the Wisconsin \Ha\
Mapper (WHAM) northern sky survey (Haffner et al.\ 2003) for cloud~g1 and the
Cohen Stream field IVC. We will describe these data when discussing each cloud.
\par Since we know the cloud distances, we can use the \Ha\ data to estimate the
amount of ionized hydrogen in our clouds. There are two ways to do this. In both
cases we assume that $N$(H$^+$) is constant, which is justified if the
ionization is due to photons from the Galactic disk: hydrogen ionization
continues until all Lyman continuum photons have been soaked up. Then, we can
first assume that the ionized hydrogen is distributed throughout the cloud, in
which case the cloud thickness we use to convert the emission measure to a
density is similar to its width on the sky. We can also assume that all the
ionized hydrogen sits on the surface of the cloud, and the cloud's density is
constant, i.e.\ we use $n$(H$^+$)=$n$(\HI) and calculate the emission length
from the emission measure. Both these methods are used to derive the numbers
presented in Table~4.

%%%%%%%%%%%%%%%%%%%%%%%%%%%%%%%%%%%%%%%%%%%%%%%%%%%%%%%%%%%%%%%%%%%%%%%%%%%%%%%%
%%%%%%%%%%%%%%%%%%%%%%%%%%%%%%%%%%%%%%%%%%%%%%%%%%%%%%%%%%%%%%%%%%%%%%%%%%%%%%%%
%%%%%%%%%%%%%%%%%%%%%%%%%%%%%%%%%%%%%%%%%%%%%%%%%%%%%%%%%%%%%%%%%%%%%%%%%%%%%%%%

\section{Results}
\subsection{Cohen Stream}
\subsubsection{Distance}
\par The first cloud we observed is part of the so-called ``Cohen Stream'',
discovered by Cohen (1981) (although he only observed the part of the stream
that lies to the east of our stars). This HVC is a rather linear stream of gas
that runs from ($l$,$b$)$\sim$(175\deg,$-$30\deg) to
($l$,$b$)$\sim$(140\deg,$-$52\deg), with velocities of \vlsr$\sim$$-$110~\kms.
It lies in a part of the sky where the maximum velocity expected from
differential galactic rotation is $-$15~\kms. A partial map of the cloud is
given in Fig.~\FCS. On its eastern end ($l$$>$156\deg) it overlaps with a bright
HVC at \vlsr=$-$280~\kms, also discovered by Cohen (1981), and named WW\,507 by
Wakker \& van Woerden (1991). It is still unclear whether these two HVCs are
related or whether they are a chance coincidence. Unfortunately, the footprint
of the SDSS does not overlap the WW\,507 cloud, so we do not (yet) know distant
stars in its direction. We return to this point in Sect.~\Sresult.1.5.
\par We obtained VLT-UVES spectra for nine stars and one QSO in the direction of
the Cohen Stream, shown in Fig.~\FsCS. Seven of these objects lie in the
direction of the small core at the end of the Stream (cloud WW\,516 in Wakker \&
van Woerden 1991), three lie projected onto its main body (cloud WW\,517). As we
discuss below, and as can be seen in Fig.~\FsCS\ and Table~3, we detect
high-velocity interstellar \CaII\ H and K absorption toward the QSO
SDSS\,J014631.99+133506.3 as well as the K line toward the 11.2$\pm$1.0~kpc
distant RR\,Lyrae star SDSS\,J015133.91+141105.2. The most distant star toward
which we do not detect the HVC at a significant level is the BHB star
2MASSJ014936.48+143914.63 at 4.6$\pm$0.8~kpc. \CaII\ is also not detected toward
SDSS\,014843.61+130411.5 ($D$=5.0$\pm$1.2~kpc) and SDSS\,020033.50+141154.0
($D$=8.7$\pm$1.3~kpc). The first of these is not significant, although just
barely so, while for the latter the ratio expected/observed equivalent width is
only 1. Taking into account the errors on the star's distances, these results
imply a distance bracket of 5.0 to 11.7~kpc for the Cohen Stream. If we relax
the criterion for a significant non-detection, we could improve the lower limit
to 5.6~kpc.

%%%%%%%%%%%%%%%%%%%%%%%%%%%%%%%%%%%%%%%%%%%%%%%%%%%%%%%%%%%%%%%%%%%%%%%%%%%%%%%%

\subsubsection{Discussion of stars}
\par 1) SDSS\,J015133.91+141105.2 is the star that sets the upper limit on the
distance. The HVC K line is clearly seen in absorption in its spectrum, with an
equivalent width of 17$\pm$3~\mA. This low-metallicity ([Fe/H]=$-$1.73$\pm$0.3)
AV star shows relatively few stellar lines, and with its velocity of
$-$176~\kms, the stellar \CaII\ and \FeI$\lambda$3931.409 lines are shifted well
away from the HVC. The expected equivalent width of the H line is 10~\mA, far
below the 34~\mA\ detection limit. This detection limit is rather high because
the stellar \He\ line substantially reduces the star's flux at the velocity of
the HVC (13.2\tdex{-16} \fu\ at \CaII~K, 3.5\tdex{-16} \fu\ at \CaII~H). Thus,
we cannot use the H-line to confirm the K-line detection. However, we have seven
exposures of this star, taken on August 10, September 7 (2 exposures) and
September 23 (4 exposures one hour apart). These show the stellar velocity
varying over time:  $v$(*)=$-$194$\pm$3~\kms\ on August 10, $-$138~\kms on
September 7, and $-$178, $-$193, $-$193 and $-$190~\kms\ on September 23. From
these spectra it is clear that the feature at \vlsr=$-$104~\kms\ is
interstellar, as it stays at the same position, whereas the stellar lines vary.
In order to have a clean continuum against which the HVC absorption can be
measured, we combined only the first and last exposure, which have the cleanest
separation between stellar and interstellar absorption.
\par 2) Toward the QSO SDSS\,J014631.99+133506.3 the high-velocity cloud's K and
H equivalent widths are 17$\pm$3 and 11$\pm$3~\mA, respectively. The \cahratio\
implied by the K-line is about half the value found toward the star
SDSS\,J15133.91+141105.2. They differ at the 2$\sigma$ level, but this can
easily be explained by the large uncertainty in the value of $N$(\HI) toward the
two targets. On the other hand, the equivalent widths of the K and H lines
seen toward the QSO are in the expected ratio of about two, giving confidence in
the detections.
\par 3) Using the method described in Sect.~\Smethod, we predict K and H
equivalent widths of 17 and 8~\mA\ toward the 30.4$\pm$3.9~kpc distant BHB star
SDSS\,J015735.65+135254.2. In its spectrum we find a possible HVC K absorption
of 10$\pm$3~\mA. Since it is just a 3$\sigma$ feature we do not claim this as a
clear detection of the HVC, but the feature is compatible with the expectations.
\par 4) Toward the star 2MASSJ\,01449.36.48+143914.6 ($D$=4.6$\pm$0.8~kpc) the
ratio of expected \CaII\ K equivalent width (21~\mA) to the observed error
(2.9~\mA) is 7.2, which implies that this star sets a lower distance limit to
the distance of the HVC.
\par 5) The HVC is also not seen in absorption toward the other six stars with
$D$$<$10~kpc. Toward most of these stars the \HI\ column density is low, and
only the non-detections toward the four nearest stars are significant.

\subsubsection{\cahratio}
\par The \cahratio\ that we find for the Cohen Stream is on the low end of those
found for HVCs: 11.7$\pm$2.9\tdex{-9} toward SDSS\,J015133.91+141105.2 and
6.8$\pm$2.0\tdex{-9} toward SDSS\,J014631.99+133506.3. The \cahratios\ predicted
by the Wakker \& Mathis (2000) correlation are 27\tdex{-9} and 17\tdex{-9},
respectively. The ratio of observed to expected values thus is 0.42$\pm$0.05,
i.e.\ the Cohen Stream appears to be relatively underabundant by a factor of
$\sim$2.4 in gaseous ionized calcium. However, until we have a measurement of
the metallicity of the Cohen Stream, it will remain unclear whether this
underabundance is due to low metallicity, unusually high calcium ionization or
high depletion of calcium onto dust, or even small-scale structure in the \HI,
which affects our estimate of $N$(\HI).

\subsubsection{Location in the Milky Way}
\par Previous assessments of the distance of the Cohen Stream were summarized by
Wakker (2001): Kemp et al.\ (1994) and Tamanaha (1996) observed \MgII, \CaII\
and \NaI\ in bright stars in the direction of the Cohen Stream, at distances up
to 300~pc, but no absorption was found. Based on a morphological study of the
high-, intermediate- and low-velocity gas, Tamanaha (1995) argued that the Cohen
Stream is the remnant of a collision between the WW\,507 cloud and low-velocity
gas, which would place it within about 1~kpc from the disk. Our distance bracket
excludes such a simple interaction model, as the cloud is clearly more distant
than 9.2~kpc.
\par It is now clear that this cloud is rather far below the Galactic plane,
having $z$ between 3.7 and 8.6~kpc (see Table~4). It is located at similar
longitude, but on the opposite side of the plane and further inward than the
Galactic warp (which is at galactocentric radii between about 17 and 22~kpc --
Burton \& te Lintel Hekkert 1986).

\subsubsection{Ionized hydrogen}
\par Weiner et al.\ (2001) reported \Ha\ emission from two directions in the
Anti Center region. One of these is at ($l$,$b$)= (164\deg,$-$46\deg) and
samples the Cohen Stream (Weiner, priv.\ comm.). The \Ha\ intensity ($I$(\Ha))
is 0.07$\pm$0.01~Rayleigh. [One Rayleigh (R) is \dex6/4$\pi$
photons~\cmm2~s$^{-1}$~sr$^{-1}$; the emission measure (EM=$\int n_e^2 L$) is
related to this as EM=2.75 $T_4^{0.924}$ I(\Ha), where $T_4$ is the gas
temperature in units of \dex4~K].
\par The H$^+$ column density we derive from the observed \Ha\ emission is on
the order of \dex{19}~\cmm2, independent of most assumptions. The H$^+$ mass
is found to be similar to the \HI\ mass, which is a few \dex5~\Msun. If we
assume the H$^+$ is located in a layer around the neutral cloud, the implied
skin thickness is 10 to 25\% of the total cloud thickness.
\par We noted earlier that \Ha\ emission was also detected from the
higher-velocity cloud near the Cohen Stream (Kutyrev \& Reynolds 1989; Weiner et
al.\ 2001), with an intensity of $\sim$0.08~R. This cloud is centered around
$l$=168\deg, $b$=$-$46\deg, \vlsr=$-$280~\kms. Using the Bland-Hawthorn \&
Maloney (1999) model of the Galactic ionizing radiation field, the \Ha\
brightness of this cloud would imply a distance on the order of 15~kpc. If such
an interpretation for the \Ha\ emission can be maintained, this would suggest
that this cloud and the Cohen Stream are close together in space, even though
their velocities differ by 170~\kms.

%%%%%%%%%%%%%%%%%%%%%%%%%%%%%%%%%%%%%%%%%%%%%%%%%%%%%%%%%%%%%%%%%%%%%%%%%%%%%%%%
%%%%%%%%%%%%%%%%%%%%%%%%%%%%%%%%%%%%%%%%%%%%%%%%%%%%%%%%%%%%%%%%%%%%%%%%%%%%%%%%

\subsection{High-z Perseus Arm IVC}
\subsubsection{Distance}
\par In the sightlines to the Cohen Stream an IVC is also visible in the \HI\
spectra, with velocities of $\sim$$-$40 to $-$60~\kms. This IVC is part of an
extended region of intermediate-velocity gas, a map of which can be found in
Wakker (2001; Fig.~18), who named this gas ``IV-south''. Figure~\FPA\ shows the
\HI\ with \vlsr\ between $-$80 and $-$30~\kms. In the general vicinity of this
IVC differential galactic rotation contributes substantially to the observed
radial velocity (up to $-$15~\kms\ at $b$=$-$50\deg, up to $-$45~\kms\ at
$b$=$-$25\deg). Therefore, we present in Fig.~\FPAdev\ the \HI\ column density
for deviation velocities between $-$60 and $-$30~\kms, i.e.\ gas parcels whose
velocities are 30 to 60~\kms\ more negative from the maximum radial velocity
expected from differential galactic rotation. This clearly shows the IVC
centered near ($l$,$b$)=(145\deg,$-$42\deg). The stellar spectra for targets
projected onto the IVC can be seen in Fig.~\FsCS. The stream running from
($l$,$b$)$\sim$(130\deg,$-$60\deg) to ($l$,$b$)$\sim$(100\deg,$-$47\deg) was
termed the ``Pegasus-Pisces Arch'' or PP-Arch by Wakker (2001), and its distance
is $<$2.7~kpc.
\par Intermediate-velocity absorption is clearly detected toward the QSO and in
the spectra of all stars more distant than 2.1~kpc, but not toward PG\,0142+148
($D$=1.2$\pm$0.5kpc), where the ratio EW(expected)/$\sigma$(EW) is 12.7. Taking
into account the errors on the stellar distances, the implied distance bracket
is 1.0 to 2.7~kpc.

\subsubsection{\cahratio}
\par For this IVC the observed \cahratios\ range from 10$\pm$3\tdex{-9} toward\\
2MASS\,J014936.48+143914.6 to 30$\pm$9 toward SDSS\,J015735.65+135254.2. Much of
this range might be due to the uncertainty in the \HI\ column densities toward
the stars. The ratio of observed to predicted \cahratio\ is on average
0.62$\pm$0.11, ranging from 0.41 to 0.75. Thus, like the Cohen Stream, this gas
also seems to be deficient in ionized calcium compared to the ratio
$N$(\CaII)/$N$(\HI) for HVCs in general, but we do not have enough information
to determine the cause of this.

\subsubsection{Location in the Milky Way}
\par The IVC lies below the Perseus Arm, the next spiral arm out from the Local
arm. In the Galactic plane, the Perseus spiral arm has velocities of about
$-$50~\kms. The distance to the Perseus arm is about 2.5$\pm$0.5~kpc (see
Reynolds et al.\ 1995 and references therein). Haffner et al.\ (2005) studied
the \Ha\ emission from the high-z extension of the Perseus Arm and find it has a
scaleheight of about 1~kpc. Associated gas at $b$=$-$45\deg\ would be at a
distance of 3.6$\pm$0.7~kpc and lie 2.5$\pm$0.5~kpc below the Galactic plane. A
$z$-height of 1~kpc implies a distance of 1.6~kpc, in which case the cloud would
be located in the interarm region. Gas near the upper distance limit of 2.7~kpc
would have $z$=1.9~kpc. So, the IVC we see is likely associated with the front
side of the high-z extension of the Perseus Arm.
\par For this direction, the LSR velocity predicted from differential galactic
rotation is $\sim$$-$25~\kms, suggesting that the IVC has a peculiar velocity
toward the plane of between 15 and 35~\kms. The derived distance bracket is not
very precise, but it is clearly compatible with the idea that this IVC is
associated with the Perseus Arm. This could be confirmed by obtaining spectra
with high S/N ratio of additional stars in the 1 to 3~kpc range.
\par To estimate the cloud's \HI\ mass we integrated the column densities
between $l$$\sim$139\deg\ to 155\deg, $b$$\sim$$-$47\deg\ to $-$40\deg. The
cloud's total mass is found to be in the range 0.2--1.4\tdex4~\Msun. We note
that this is of the same order of magnitude as the ``Intermediate-Velocity
Arch'' (IV-Arch), the largest northern IVC (see Kuntz \& Danly 1996; Wakker
2001, 2004), which also is moving toward the Galactic plane with a velocity of
$\sim$40~\kms. It is also similar to the mass of the ``Low-Latitude
Intermediate-Velocity Arch'' (LLIV Arch), an IVC that is 0.9--1.8~kpc distant
with a mass of 1.5\tdex5~\Msun. The latter IVC appears to be in the interarm
region. The IVC we detected in absorption may be a cloud similar to the IV and
LLIV Arches, but associated with the Perseus Arm, rather than with the Local
spiral arm.

\subsubsection{Ionized hydrogen}
\par In the WHAM \Ha\ survey (Haffner et al.\ 2005), there is no detectable \Ha\
emission from the area covered by this cloud, which corresponds to a 3$\sigma$
upper limit of $I$(\Ha)$<$0.15~R, or an emission measure $<$0.07~\cmm6\,pc. The
implied H$^+$ column density limit of $\sim$\dex{19}~\cmm2\ is compatible with
the values found for the other clouds. so the non-detection does not imply a
lack of ionized hydrogen for this object.
\par The non-detection does imply that for this cloud the mass of ionized gas is
smaller than the neutral hydrogen mass. If the ionized gas is on a skin, this
skin is relatively thin, being at most 10\% of the cloud's thickness.

%%%%%%%%%%%%%%%%%%%%%%%%%%%%%%%%%%%%%%%%%%%%%%%%%%%%%%%%%%%%%%%%%%%%%%%%%%%%%%%%
%%%%%%%%%%%%%%%%%%%%%%%%%%%%%%%%%%%%%%%%%%%%%%%%%%%%%%%%%%%%%%%%%%%%%%%%%%%%%%%%

\subsection{HVC complex~GCP or HVC\,40$-$15+100}
\subsubsection{Distance}
\par The second HVC in our sample lies near $l$=40\deg, $b$=$-$15\deg\ and has
\vlsr$\sim$90~\kms. It has been referred to as the ``Smith Cloud'', after Smith
(1963), who discovered it, or as HVC\,40$-$15+100, or as complex~GCP (Wakker \&
van Woerden 1991). Wakker (2001) could only set a lower distance limit of
0.3~kpc, from the non-detection of \CaII\ absorption toward some nearby stars.
\par We observed eight stars projected on complex~GCP (see Fig.~\FGP), six of
which were found from the 2MASS survey; the other two are RR\,Lyraes from the
Kukarkin (1970) catalogue. The \CaII\ absorption and \HI\ emission spectra are
shown in Fig.~\FsGP. We detect the HVC in the spectrum of V1084\,Aql, located at
14.5$\pm$1.3~kpc, but not toward V1172\,Aql at 10.5$\pm$1.3~kpc, yielding a
distance bracket of 9.8 to 15.1~kpc.

\subsubsection{Discussion of stars}
\par 1) Figure~\FsGP\ shows the HVC absorption in the V1084\,Aql K-line, which
has an equivalent width of 51$\pm$3~\mA. We caught this low-metallicity
([Fe/H]=$-$1.22$\pm$0.25) RR\,Lyrae star at a temperature of about 7000~K, and
the spectrum shows relatively few stellar lines. The \CaII\ H-line detection is
less clear (37$\pm$10~\mA), not only because it is confused by the stellar \He\
line, but also because the \He\ lines makes the flux much lower at the HVC's
velocity (8.5\tdex{-16} \fu\ at K, 1.5\tdex{-16} \fu\ at H).
\par 2) The HVC is absent in the spectrum of V1172\,Aql ($D$=10.5$\pm$1.4~kpc),
and the ratio expected/error is 15.5 for \CaII\ K, 7.1 for \CaII\ H. This star
thus yields a lower limit for the HVC's distance. We note that the spectrum of
V1172\,Aql shows weak ($\sim$25~\mA) wiggles at the velocity of the HVC in both
the K and the H line, but these features don't match up as interstellar \CaII\
-- they are not at exactly the same velocity, and they are not in the ratio 2:1.
\par 3) One other star has a large distance (2MASS\,J195922.75+000300.0 at
9.0$\pm$3.7~kpc), but it has low UV flux as it is a cool star ($T_{\rm
eff}$=5750~K), so the non-detection is not significant.
\par 4) The remaining four stars all have distances less than 2~kpc. The
non-detections of interstellar \CaII\ K toward 2MASS\,J195927.29+000822.3 and
2MASS\,J195912.00$-$002645.3 are significant, but for
2MASS\,J195741.61$-$004009.7 and 2MASSJ\,195925.49$-$000519.1 this is not the
case. Considering the distance bracket we have set, these non-detections are not
unexpected since these stars are relatively nearby.
\par 5) Finally, one star in our sample is at about the same distance as
V1084\,Aql:\\2MASS\,J195823.36$-$002719.0, at $D$=13.1$\pm$3.2~kpc. The 2MASS
catalogue gives a $J$ magnitude of 15.1 for this star, but our photometry
(Wilhelm et al.\ 2007) yields $V$=16.9. Since we expected the star to be bright
(and originally estimated a distance of 5.8~kpc), we only requested a short
exposure time (7200~sec). Even less data was actually obtained (2665~sec), so
the spectrum has an S/N ratio of only 5 at the location of the interstellar K
line. Therefore, the non-detection is not significant (61~\mA\ expected,
3$\sigma$ upper limit 67~mA).

\subsubsection{\cahratio}
\par Weighing the K line twice as much as the H line, the average \cahratio\ in
complex~GCP that we find from the K and H line detected toward V1084\,Aql is
6.2$\pm$1.7\tdex{-9}. If we estimate EW(K) from the Wakker \& Mathis (2000)
relation between log~$R$(\CaII) and log~$N$(\HI), we predict
$R$(\CaII)$\sim$6\tdex{-9} for complex~GCP. Thus, the \cahratio\ is exactly what
is expected.

\subsubsection{Location in the Milky Way}
\par We place complex~GCP at a distance of 9.8 to 15.1~kpc, which puts it at a
galactocentric radius between 6.2 and 9.7~kpc, and 2.5 to 3.9~kpc below the
plane. Its 2\deg$\times$8\deg\ size on the sky implies a linear size of about
0.5$\times$1.8~kpc. The \HI\ mass is about \dex6~\Msun.
\par Bland-Hawthorn et al.\ (1998) suggested that this HVC is related to the
Sagittarius dwarf, at a distance of 26$\pm$4\,kpc ($R$=20~kpc, $z$=$-$7~kpc). We
find that instead of lying on the outskirts of the Milky Way, this HVC is near
the solar circle, only halfway to the Sagittarius dwarf. Only a detailed model
of the dwarf galaxy's orbit can tell whether the HVC could be related after all.
\par In the direction of complex~GCP differential galactic rotation can explain
a velocity of up to +75~\kms, which occurs at the tangent point ($D$=7~kpc, at
galactocentric radius 5.5~kpc, 2.0~kpc below the plane). At distances of 9.8 and
15.1~kpc, gas in differential rotation has an LSR velocity of +50 and
$-$17~\kms, respectively. It is therefore clear that complex~GCP is below the
Galactic plane and moving ahead of the gas in the plane with a velocity of 40 to
105~\kms. Heroux et al.\ (in preparation) mapped this HVC using the new
Green-Bank Telescope. They are working on a model of its orbit, including drag
forces to explain the stretching that is obvious in the sky map (Fig.~\FGP).

\subsubsection{Ionized hydrogen}
\par Bland-Hawthorn et al.\ (1998) based their suggested association with the
Sagittarius dwarf on the detected \Ha\ emission from this HVC, with intensities
0.24~R at ($l$,$b$)=(40\deg,$-$15\deg) and 0.30~R at
($l$,$b$)=(41\deg,$-$14\deg). If we assume that the cloud's thickness is
comparable to the width of its short dimension (0.3--0.5~kpc), these emission
measures imply an electron density $n_e$ of 0.04--0.03~\cmm3, about 25\% of the
implied \HI\ density of 0.16--0.10~\cmm3. In that case the H$^+$ mass is similar
to the cloud's \HI\ mass. If on the other hand we assume the H$^+$ occurs around
the outside of the cloud, the ionized mass is about a third of the neutral mass,
while the skin is about 10\% of the thickness of the cloud. In either case it is
clear that this cloud must have a substantial ionized component. Now that we
know its location, a full \Ha\ map would clearly help understand the Galactic
ionizing radiation field.

%%%%%%%%%%%%%%%%%%%%%%%%%%%%%%%%%%%%%%%%%%%%%%%%%%%%%%%%%%%%%%%%%%%%%%%%%%%%%%%%
%%%%%%%%%%%%%%%%%%%%%%%%%%%%%%%%%%%%%%%%%%%%%%%%%%%%%%%%%%%%%%%%%%%%%%%%%%%%%%%%

\subsection{Cloud~g1}
\subsubsection{Distance}
\par The last cloud in our sample is an IVC near ($l$,$b$)=(67\deg,$-$27\deg),
at \vlsr$\sim$+70~\kms\ (Fig.~\Fgp). It is one of a set of faint
positive-velocity IVCs in the region $l$=30\deg\ to 70\deg, $b$=$-$45\deg\ to
$-$20\deg\ that Wakker (2001) termed ``complex~gp''. For the present project we
have designated some of the cores in this complex as g1 through g5, with the IVC
toward which we observed stars termed ``cloud~g1''. It is, a-priori, unclear
whether these IVCs are related to complex~GCP, which lies near the IVCs, though
at much higher positive velocity.
\par Many stars projected onto cloud~g1 have previously been observed. It was
detected in \CaII\ and \NaI\ absorption toward several stars in M\,15
($D$=10~kpc; Cohen 1979; Songaila \& York 1981; Kennedy et al.\ 1998; Lehner et
al.\ 1999). Intermediate-velocity absorption at the same velocity as cloud~g1
was also seen toward the star HD\,203664 ($D$=4.3$\pm$0.4~kpc; Albert et al.\
1993; Little et al.\ 1994; Ryans et al.\ 1996), which lies three degrees west of
the main cloud core (see Fig.~\Fgp). Absorption is not detected toward
HD\,203699 ($D$=0.8$\pm$0.1~kpc; Kennedy et al.\ 1998), nor toward several A and
F stars at distances between 0.2 and 1.4~kpc (Smoker et al.\ 2001). The
non-detections in the latter paper are usually not significant by our criteria,
except maybe for their star \#7, an A2V star at a distance of 1.4~kpc. Together,
these detections and non-detections bracketed the distance of g1 as 0.7 (maybe
1.4) to 4.5~kpc, with the caveat that these stars do not lie near the brighter
part of the cloud, but probe the diffuse distribution of gas with similar
velocities.
\par Smoker et al.\ (2002) made a detailed study of the part the cloud that lies
in front of M\,15. They combined high angular resolution
(1\arcmin$\times$2\arcmin) \HI\ (Westerbork+Arecibo) data with WHAM \Ha\
profiles and an IRAS map of the region. They found that the \HI\ has structure
on scales of 5 arcmin (1.5~$D$[kpc]~pc), and that the gas is warm ($\sim$500~K).
\par We observed seven stars projected onto the main core of cloud~g1, with
distances ranging from 1.2 to 8.2~kpc. Since these stars are relatively nearby,
the spectra have relatively high S/N ratios. Absorption is seen toward three
stars: 2MASS\,J213327.04+133026.7 ($D$=8.2$\pm$0.9~kpc), FW\,Peg
($D$=3.8$\pm$0.8~kpc) and BS\,17578-0016 ($D$=3.6$\pm$0.5) while significant
non-detections are obtained toward the other four stars (ratios of expected EW
to upper limit of 25 to 60 for \CaII\ K). The most distant of these is
PG\,2134+125 at 2.2$\pm$0.8~kpc. Using the 1$\sigma$ range allowed by the
stellar distances, the distance of cloud~g1 is then bracketed as 1.8 to
3.8~kpc. This is a range of 2~kpc, tightening the previous 3.8~kpc wide
bracket. Unfortunately, the current distance bracket still means that the
derived cloud parameters will have an uncertainty of a factor 2 to 4 (depending
on how they scale with distance).

\subsubsection{Discussion of stars}
\par 1) As Fig.~\Fsgp\ shows, the four non-detections toward BS17578-0015,
2MASS\,J213520.04+133045.0 2MASS\,J213451.58+134017.5, and PG\,2134+125 are very
clear. The spectra show no stellar lines near the HVC velocity, and the stellar
\CaII~K, H and \He\ lines are not interfering.
\par 2) The detection toward FW\,Peg is simple and certain: the \CaII~K and H
absorption lines have equivalent width ratio two, to within the errors. The
stellar velocity of $-$225~\kms\ shifts the stellar \CaII\ K nicely out of the
way. Further, we have two spectra of this star (taken 16 days apart) in which
the stellar lines clearly shift in velocity, but the lines interpreted as
interstellar do not. A few (unidentified) stellar lines can be seen toward more
positive velocities than the IVC absorption, but they do not confuse the issue.
\par 3) The detection toward 2MASS\,J213327.04+133026.7 is not as clean as the
one toward FW\,Peg, but the line is obviously interstellar, since the spectrum
shows no stellar lines other than \CaII\ K, H and \He. However, the \He\ line is
very broad and deep, lowering the flux near the velocity of the interstellar
absorption and making the detection of \CaII~H more noisy.
\par 4) The most problematic star is BS\,17578-0016. Both the \HI\ 21-cm profile
and the \CaII\ absorption lines show two narrow components, separated at the
velocity where the \HI\ in g1 usually peaks. This actually gives confidence in
the interpretation of the absorption as associated with g1, but it makes the
measurements more difficult. In Table~3 we list the absorption as two separate
components.

\subsubsection{\cahratio}
\par As summarized by Wakker (2001), Lehner et al.\ (1999) detected \CaII\ in
cloud~g1 toward 13 stars in M\,15. Using values of $N$(\HI) between 42 and
54\tdex{18}~\cmm2 for these stars gives \cahratios\ between 14\tdex{-9} and
79\tdex{-9}, averaging to 35\tdex{-9}, although individual measurements have
large errors. Toward HD\,203664 Ryans et al.\ (1996) derived a \cahratio\ of
450$\pm$40\tdex{-9} (for $N$(\HI)=2.2\tdex{18}~\cmm2). For the three stars
toward which we detect cloud~g1 we find values of 106$\pm$22\tdex{-9} (FW\,Peg),
63$\pm$16\dex{-9} (2MASS\,J213327.04+133026.7), 109$\pm$28\tdex{-9}
(BS\,17578-0016, stronger \HI\ component) and 510$\pm$110\tdex{-9}
(BS\,17578-0016, weaker \HI\ component). Using the Wakker \& Mathis (2000)
relation, we would predict \cahratios\ of 11\tdex{-9} for the stars in M\,15,
140\tdex{-9} toward HD\,203664, and 28\tdex{-9}, 35\tdex{-9}, 103\tdex{-9} and
225\tdex{-9} for our detections. The ratio of predicted to observed \cahratio\
thus varies from 1.1 for one of the BS\,17578-0016 components, to 3.8 for
FW\,Peg. However, the \HI\ column densities are comparatively uncertain, since
the cloud is known to have structure on much smaller scales than the 36\arcmin\
Dwingeloo beam (Smoker et al.\ 2002). So, a proper measurement of the
\cahratios\ in cloud~g1 will require \HI\ observations with much higher
resolution toward each of the stellar targets. Nevertheless, it appears that the
\cahratio\ is higher than average in cloud~g1. Further, the apparently extremely
high \cahratios\ for the weak components toward BS\,17578-0016 are in line with
the values expected from the Wakker \& Mathis (2000) relation.

\subsubsection{Location in the Milky Way}
\par The distance bracket of 1.8--3.8~kpc for cloud~g1 places it at a
galactocentric radius of $\sim$7.9~kpc and $z$=$-$0.8 to $-$1.7~kpc, moving with
$v$=55~\kms\ relative to co-rotating Galactic gas around it. The ranges in the
implied galactocentric radius and peculiar velocity are small because g1 lies
near the tangent point. Its height below the plane is typical for an IVC.
However, unlike the large Intermediate-Velocity Arch in the northern sky, this
gas is moving away from the plane with an estimated velocity of about 25~\kms.
\par The area of the core of the cloud is only about 2\deg$\times$2\deg, which
translates to a size of (60--130)$^2$~pc. The \HI\ mass is only about
1000--4000~\Msun. The neutral part of this cloud is thus clearly fairly small.

\subsubsection{Ionized hydrogen}
\par Smoker et al.\ (2002) reported \Ha\ emission from cloud~g1 toward three
directions that lie projected onto the areas with the brightest \HI\ emission.
These detections were taken from the WHAM northern sky survey (Haffner et al.\
2003). The \Ha\ intensity does not vary much (the detections are
1.30$\pm$0.04~R, 0.95$\pm$0.04~R and 0.91$\pm$0.03~R), whereas the \HI\ column
densities have a range of a factor 3 for these directions. This result supports
the contention that for many HVCs/IVCs $I$(\Ha) will be relatively constant
across the face of the cloud, and not correlated with $N$(\HI). With our
distance bracket for this cloud, the bright \Ha\ intensities imply either a
relatively large H$^+$ volume density ($\sim$0.17~\cmm3) or a large ionized
pathlength (several hundred pc). With either of our assumptions, $M$(H$^+$) is
much larger than $M$(\HI), by at factor 5 to 10. It is clear that what we call
cloud~g1 is a small neutral condensation in a much larger region of ionized gas.

%%%%%%%%%%%%%%%%%%%%%%%%%%%%%%%%%%%%%%%%%%%%%%%%%%%%%%%%%%%%%%%%%%%%%%%%%%%%%%%%

\subsection{Low-velocity gas}
\par Low-velocity \CaII\ absorption is detected toward all stars, although in a
few cases stellar lines make it impossible to measure the equivalent widths. It
is likely that the low-velocity absorption is a mixture of several nearby clouds
in the line of sight, but we have no way of separating these.
\par The \cahratios\ measured for the low-velocity gas are similar for the
different fields, averaging to 4.4$\pm$1.6\tdex{-9} for stars projected on the
Cohen Stream, 5.3$\pm$1.9\tdex{-9} for the complex~GCP stars and
5.7$\pm$2.0\tdex{-9} for stars in the cloud~g1 field. Applying the Wakker \&
Mathis (2000) relation, these values are only a factor 2--4 higher than
expected. This is remarkable, since that relation was determined using HVC and
IVC detections only, and in principle it might not be applicable to low-velocity
absorption components since these are probably a mixture of several clouds with
different column densities. Also, the column densities of the low-velocity gas
are a factor 10 to a 100 larger than the column densities in the HVCs.
\par In the direction to complex~GCP there is an absorption component at
\vlsr$\sim$30~\kms\ that is detected in every sightline. This component is well
separated in the \CaII\ absorption, but not in the \HI\ profile. To estimate
$N$(\HI) we therefore integrated the \HI\ over the same velocity range as the
\CaII\ absorption. The measured \cahratio\ is rather uncertain because the
associated \HI\ column density is very difficult to measure. Using a flat
rotation curve this gas would have a kinematical distance of 2.0~kpc
($R$=7.1~kpc, $z$=$-$0.5~kpc). This is consistent with the detection toward the
star at 1.8$\pm$0.3~kpc, but the detections toward the star that are
0.7$\pm$0.6~kpc and 1.1$\pm$0.5~kpc distant suggest that the gas is much nearer
than indicated by the simple kinematical model. At a distance of 0.7~kpc, the
expected radial velocity is only $\sim$8~\kms, so this gas deviates by about
20~\kms\ from its expected velocity.

%%%%%%%%%%%%%%%%%%%%%%%%%%%%%%%%%%%%%%%%%%%%%%%%%%%%%%%%%%%%%%%%%%%%%%%%%%%%%%%%
%%%%%%%%%%%%%%%%%%%%%%%%%%%%%%%%%%%%%%%%%%%%%%%%%%%%%%%%%%%%%%%%%%%%%%%%%%%%%%%%

\section{Discussion}
\par We now present a short discussion summarizing the implications of our
results as they pertain to the mass inflow rate of external gas, to the study of
the escape of ionizing radiation from the Milky Way and to the understanding of
tidal streams around the Milky Way. We do not give a full discussion of these
items -- that will be saved for a future paper, after we have determined more
HVC and IVC distances.
\par One of the more important processes that HVCs trace is the infall of
low-metallicity materal onto the Milky Way. Wakker et al.\ (1999) found that HVC
complex~C (which has a metallicity 0.15 times solar -- Wakker et al.\ 1999;
Richter et al.\ 2001; Tripp et al.\ 2003; Sembach et al.\ 2004) contributes a
mass inflow rate of about 0.1--0.2~\Msunpyr, assuming a distance of 10~kpc. A
similar calculation for HVC complex~A shows that it contributes about
0.05~\Msunpyr. Although we do not know the metallicity of the Cohen Stream, its
location and velocity leads us to predict that it is a low-metallicity newcomer
to the Milky Way. As Table~4 shows, it contributes about 0.01~\Msunpyr.
\par There are a number of other HVCs that may represent low-metallicity
infalling gas. However, we do not have direct data to prove this. These are the
HVCs in complexes H, ACHV, ACVHV, GN, P (see Wakker \& van Woerden 1991 for the
definition). If all of these objects were at a distance of 10~kpc, they would
represent an inflow rate of 0.4~\Msunpyr. The Cohen Stream is one of the larger,
but not one of the brighter HVCs in the ACHV complex. If all clouds in this
complex are 10~kpc distant, the complex as a whole corresponds to a mass flow
rate of 0.1~\Msunpyr.
\par Putman et al.\ (2004) proposed that most, if not all, of the HVCs in the
Anti Center region are in the orbit of the Sagittarius dwarf, and they suggested
that this gas was stripped from that galaxy 200 to 300~Myr ago as it interacted
with the gas at the far outskirts of the Galaxy. This would imply a distance on
the order of 40~kpc for the Anti Center HVCs. We cannot directly exclude that
the higher velocity HVC in the region (\vlsr$\sim$$-$280~\kms) is this distant,
but clearly the Cohen Stream is much closer. Moreover, the intensity of the \Ha\
emission from the $-$280~\kms\ cloud is similar to that of the Cohen Stream,
which argues that their distances are similar.
\par We find that the IVC near ($l$,$b$)=(140\deg,$-$40\deg) could well be a
cloud associated with the Perseus Arm. We do not have an accurate distance, but
even so we find that its properties are comparable to those of the two large
northern IVCs, the IV-Arch and the LLIV-Arch. Both of these have near-solar
metallicity, a mass on the order of \dex5~\Msun, and each represent an inflow of
about 5\tdex{-3}~\Msunpyr (Wakker 2004). Thus, if we were living in the Perseus
Arm, this IVC would show up as a fairly large solar-metallicity infalling cloud
near the Galactic south pole.
\par The g1 cloud has properties just like what is expected for gas taking part
in the outflow phase of a Galactic Fountain (see Houck \& Bregman 1991). As the
hot gas flows out it starts condensing a few kpc up, before it reaches its
maximum height (cloud~g1 is at $z$$\sim$1~kpc, moving out with 25~\kms). The
process of ejecting the gas destroys much of the dust, leading to relatively
high gaseous abundances (cloud~g1 has a larger than average \cahratio). But most
of the gas is still ionized ($M$(H$^+$)$\sim$5--10 $M$(\HI) for cloud~g1). The
only thing we still have not measured directly is the metallicity of cloud~g1,
which will require spectra in the 1000--1300~\AA\ spectra range for UV-bright
stars projected onto it. We expect the metallicity to be solar.
\par Finally, we note that some of the conclusions above can be confirmed by
measuring the cloud metallicities. This can be done by determining the abundance
of \OI, \SII\ and other ions in spectra of UV-bright targets located behind our
clouds. A check of the FUSE and HST archives shows that no useful data exists at
present. Targets behind the clouds have been observed, but their spectra have
too low resolution, there are too many stellar lines, or they are too faint to
provide enough background flux.

%%%%%%%%%%%%%%%%%%%%%%%%%%%%%%%%%%%%%%%%%%%%%%%%%%%%%%%%%%%%%%%%%%%%%%%%%%%%%%%%
%%%%%%%%%%%%%%%%%%%%%%%%%%%%%%%%%%%%%%%%%%%%%%%%%%%%%%%%%%%%%%%%%%%%%%%%%%%%%%%%

\section{Conclusions}
\par We report on VLT observations of 24 stars and one QSO projected on four
high-velocity clouds (HVC). Using these data we derive distances to these
objects. We combine \HI\ and \Ha\ data to estimate the total cloud (neutral and
ionized) gas masses.
\par (1) We derive a distance bracket of 5.0 to 11.7~kpc for Cohen Stream, a
cloud that is the most-negative-latitude large cloud in the Anti-Center complex
of HVCs. This places the cloud between 3.7 and 8.6~kpc below the Galactic plane.
The stream runs from ($l$,$b$)$\sim$(175\deg,$-$30\deg) to
($l$,$b$)$\sim$(140\deg,$-$52\deg), and has velocities of
\vlsr$\sim$$-$110~\kms. \CaII\ K and H absorption is detected toward the QSO
SDSS\,J014631.99+133506.3 with equivalent widths of 17$\pm$3 and 11$\pm$3~\mA,
respectively. We also find a 17$\pm$3~\mA\ \CaII\ K absorption in the spectrum
of the 11.2$\pm$1.0~kpc distant RR\,Lyrae star SDSS\,J015133.91+141105.2. The
lower limit is derived from the significant non-detection of high-velocity
\CaII\ toward the BHB star 2MASS\,J014936.48+143914.6 (21~\mA\ line expected,
3$\sigma$ upper limit 9~\mA).
\par (2) The distance bracket for the Cohen Stream implies a total neutral
hydrogen mass of 2.5 to 3.9\tdex{5}~\Msun\ for the cloud, and an associated mass
inflow rate on the order of 0.004~\Msunpyr. \Ha\ emission from the Cohen Stream
has been detected at one position and its intensity suggests an amount of
ionized gas that is comparable to the amount of neutral gas. The complex of HVCs
that the Cohen Stream is a part of would represent a total inflow of
0.1~\Msunpyr\ if all clouds in the complex are at similar distances. A direct
measurement of the metallicity of this complex is still lacking, so the
interpretation of the complex as infalling low-metallicity gas is not confirmed.
\par (3) We find a rough distance bracket of 1.0 to 2.7~kpc for
intermediate-velocity gas in the direction of the Cohen Stream. This suggests
that this gas is part of a high-$z$ extension of the Perseus Arm, similar to the
more well-studied northern intermediate-velocity clouds. Like these clouds the
IVC may be part of a Galactic Fountain flow, but it is associated with the
Perseus Arm rather than with the Local Arm.
\par (4) We determine the distance to HVC complex~GCP as 9.8 to 15.1~kpc, based
on the detection of \CaII\ K absorption (51$\pm$3~\mA) in the spectrum of the
RR\,Lyrae star V1084\,Aql ($D$=14.5$\pm$1.3~kpc), and the significant
non-detection (3$\sigma$ limit 11~\mA, expected 59~\mA) toward the RR\,Lyrae
V1172\,Aql ($D$=10.5$\pm$1.4~kpc).
\par (5) The distance bracket for complex~GCP implies the cloud is at a
galactocentric radius of 6.2 to 9.7~kpc, lies 2.5 to 3.9~kpc below the plane and
has a neutral gas mass of $\sim$\dex6~\Msun. \Ha\ emission is detected from two
positions in the cloud and implies an ionized component that is 40 to 120\% as
large as the neutral component. At its location the cloud is moving with a
velocity of 35--75~\kms\ relative to gas taking part in differential galactic
rotation, and it is moving away from the Galactic plane at about 10--20~\kms.
\par (6) For the IVC cloud~g1 we derive a distance bracket of 1.8 to 3.8~kpc,
detecting the associated absorption in the spectra of three stars
2MASS\,J213327.04+133026.7 ($D$=8.2$\pm$0.9~kpc), FW\,Peg ($D$=3.6$\pm$0.8~kpc)
and BS\,17578-0016 ($D$=3.6$\pm$0.5), but not in four nearer stars. This places
the IVC 0.8 to 1.7~kpc below the Galactic disk, and implies that it has a
peculiar velocity of about 55~\kms. It is a small cloud, with a mass of no more
than \dex4~\Msun, and an area of (60--130)$^2$~pc$^2$. It is moving away from
the Galactic plane with a velocity of about 20~\kms. The relatively bright \Ha\
emission coming from this cloud (Smoker et al.\ 2002) implies that there is a
factor 5 to 10 more ionized gas than neutral gas. The most likely interpretation
for this object appears to be that it is condensing in an upward Fountain flow.
\par (7) The \cahratios\ that we derive are all in line with the relation
between $R$(\CaII) and $N$(\HI) that was found by Wakker \& Mathis (2000),
except that each cloud appears to have a different scaling factor. In
particular, the Cohen Stream detections are a factor 0.44$\pm$0.5 below the
average relation, for the Perseus Arm gas the scaling factor is 0.62$\pm$0.11,
the complex~GCP detection is on the average, while the cloud~g1 detections fall
a factor 2.2$\pm$1.0 above the average.
\par (8) We have set new distance brackets to four HVCs/IVCs, which trace at
least two of the different origins of HVCs: inflow of new material (the Cohen
Stream), the Galactic Fountain (cloud~g1 and the Perseus Arm IVC). The origin of
the fourth cloud (complex~GCP) is less clear, but with our distance estimate
detailed modeling is now possible. In the future we hope to determine more
distances to HVCs/IVCs, and we should eventually be able to measure the inflow
rate of accreting material as well as the rate of circulation of gas between
disk and halo.

\acknowledgements
\par B.P.W., D.G.Y, R.W. and T.C.B. acknowledge support from grant AST-06-07154
awarded by the US National Science Foundation. T.C.B. also acknowledges NSF
grant PHY-02-16783; Physics Frontier Center/Joint Institute for Nuclear
Astrophysics (JINA).
\par We thank Ben McCall for help in observing logistics support, the staff of
Apache Point Observatory for excellent observing support and Chris Thom for
discussions.
\par Funding for the SDSS and SDSS-II has been provided by the Alfred P. Sloan
Foundation, the Participating Institutions, the National Science Foundation, the
U.S. Department of Energy, the National Aeronautics and Space Administration,
the Japanese Monbukagakusho, the Max Planck Society, and the Higher Education
Funding Council for England. The SDSS Web Site is http://www.sdss.org/. 
\par The SDSS is managed by the Astrophysical Research Consortium for the
Participating Institutions. The Participating Institutions are the American
Museum of Natural History, Astrophysical Institute Potsdam, University of Basel,
University of Cambridge, Case Western Reserve University, University of Chicago,
Drexel University, Fermilab, the Institute for Advanced Study, the Japan
Participation Group, Johns Hopkins University, the Joint Institute for Nuclear
Astrophysics, the Kavli Institute for Particle Astrophysics and Cosmology, the
Korean Scientist Group, the Chinese Academy of Sciences (LAMOST), Los Alamos
National Laboratory, the Max-Planck-Institute for Astronomy (MPIA), the
Max-Planck-Institute for Astrophysics (MPA), New Mexico State University, Ohio
State University, University of Pittsburgh, University of Portsmouth, Princeton
University, the United States Naval Observatory, and the University of
Washington.

%%%%%%%%%%%%%%%%%%%%%%%%%%%%%%%%%%%%%%%%%%%%%%%%%%%%%%%%%%%%%%%%%%%%%%%%%%%%%%%%

%%%%%%%%%%%%%%%%%%%%%%%%%%%%%%%%%%%%%%%%%%%%%%%%%%%%%%%%%%%%%%%%%%%%%%%%%%%%%%%%
%%%%%%%%%%%%%%%%%%%%%%%%%%%%%%%%%%%%%%%%%%%%%%%%%%%%%%%%%%%%%%%%%%%%%%%%%%%%%%%%

\clearpage
\def\e{$\pm$}
\def\v{\hbox to 0pt{$^{10}$\hss}}
\def\kms{km\,s$^{-1}$}
\begin{deluxetable}{lllcccclrrcrr}
\rotate
\tabletypesize{\footnotesize}
\tabcolsep=2pt \tablenum{1} \tablewidth{0pt}
\tablecolumns{13}
\tablecaption{Stellar data}
\tablehead{%
\colhead{object}         &\colhead{Tel.$^1$}      &\colhead{phot$^2$}    &\colhead{y/V/r$^3$}&\colhead{B/g$^3$} &\colhead{U/u$^3$}&\colhead{$E_{B-V}$}&
\colhead{T$_{\rm eff}^4$}&\colhead{$log~g^5$}     &\colhead{$log Z^6$}   &\colhead{type$^7$} &\colhead{$v(*)^8$}&\colhead{$D^9$}  \\
\colhead{   }            &\colhead{   }           &\colhead{   }         &\colhead{   }      &\colhead{   }     &\colhead{   }    &\colhead{   }      &
\colhead{[K]}            &\colhead{[cm\,s$^{-2}$]}&\colhead{[/$Z_\odot$]}&\colhead{}         &\colhead{[\kms]}  &\colhead{[kpc]}  \\
\colhead{(1)}            &\colhead{(2)}           &\colhead{(3)}         &\colhead{(4)}      &\colhead{(5)}     &\colhead{(6)}    &\colhead{(7)}      &
\colhead{(8)}            &\colhead{(9)}           &\colhead{(10)}        &\colhead{(11)}     &\colhead{(12)}    &\colhead{(13)}   }
\startdata
PG0142+148$^{11}$       & VLT  & Str  & 13.73 &       &       & 0.048 &   26200     & 5.10       &             & sdB    &  $-$74\e2  &  1.2\e0.5 \\ % PG0142+148
                        & APO  &      &       &       &       &       &   26200     & 5.10       &             & sdB    &  $-$67     &  1.2\e0.5 \\ % PG0142+148
2MASSJ014825.88+132305.3& VLT  & UBV  & 13.89 & 14.06 & 14.01 & 0.071 &   7927\e140 & 4.08\e0.78 & $-$1.99\e0.12 & AV     & $-$167\e2  &  2.1\e1.3 \\ % 2MASSJ014825.88+132305.3 % OK, checked
                        & APO  &      &       &       &       &       &   8120\e141 & 4.47\e0.63 & $-$1.70\e0.12 & AV     & $-$155     &  1.2\e0.7 \\ % 2MASSJ014825.88+132305.3
2MASSJ021651.34+080150.2& VLT  & 2MS  & 16.80 &       &       & 0.102 &   6250\e250 & 4.50\e0.50 & $-$1.00\e0.25 & FV     &  213\e1  &  2.2\e1.0 \\ % 2MASSJ021651.34+080150.2 % GET RON 2MS
                        & APO  &      &       &       &       &       &   6250\e250 & 2.75\e0.50 & $-$1.25\e0.25 & FHB    &  209     & 21.0\e8.6 \\ % 2MASSJ021651.34+080150.2
2MASSJ021409.31+090105.3& VLT  & UBV  & 13.55 & 13.79 & 13.95 & 0.092 &   7744\e46  & 3.47\e0.17 & $-$1.39\e0.16 & AV     &  $-$37\e1  &  3.3\e0.8 \\ % 2MASSJ021409.31+090105.3 % OK, checked
                        & APO  &      &       &       &       &       &   7646\e76  & 3.25\e0.12 & $-$1.71\e0.09 & FHB    &   $-$2     &  3.1\e0.2 \\ % 2MASSJ021409.31+090105.3
                        & APO  &      &       &       &       &       &   7749\e115 & 3.53\e0.16 & $-$1.14\e0.14 & AV     &  $-$25     &  3.3\e0.6 \\ % 2MASSJ021409.31+090105.3
2MASSJ014936.48+143914.6& VLT  & UBV  & 14.05 & 14.21 & 14.27 & 0.044 &   8113\e106 & 3.42\e0.27 & $-$0.03\e0.06 & AV     &  $-$66\e7  &  4.6\e0.8 \\ % 2MASSJ014936.48+143914.6 % OK, checked, Ron will rerun
                        & APO  &      &       &       &       &       &   7901\e233 & 3.34\e0.37 & $-$0.81\e0.28 & FHB/AV &  $-$30     &  4.0\e1.0 \\ % 2MASSJ014936.48+143914.6
SDSSJ014843.61+130411.5 & VLT  & SDSS & 14.20 & 14.59 & 15.84 & 0.084 &   5880\e112 & 3.00\e0.25 & $-$0.66\e0.11 & FV     &  $-$57\e2  &  5.0\e1.2 \\ % SDSSJ014843.61+130411.5  % OK, ugr+>UBV
                        & APO  & UBVS & 14.43 & 15.02 & 15.07 &       &   5774\e100 & 2.75\e1.07 & $-$0.89\e0.32 & FHB    &  $-$97     &  6.3\e2.5 \\ % SDSSJ014843.61+130411.5
SDSSJ020033.50+141154.0 & VLT  & SDSS & 15.48 & 15.58 & 16.89 & 0.059 &   6644\e22  & 2.00\e0.00 & $-$1.32\e0.18 & RR     &  $-$29\e3\v&  8.7\e1.3 \\ % SDSSJ020033.50+141154.0  % OK, ugr+>UBV
                        & SDSS & UBVS & 15.58 & 15.87 & 15.97 &       &   6911\e21  & 2.15\e0.09 & $-$1.73\e0.21 & RR     &    8     &           \\ % SDSSJ020033.50+141154.0
SDSSJ015133.91+141105.2 & VLT  & SDSS & 16.64 & 16.89 & 17.94 & 0.049 &   6823\e312 & 3.87\e0.37 & $-$1.90\e0.22 & RR     & $-$190\e2\v& 11.2\e1.0 \\ % SDSSJ015133.91+141105.2  % OK, ugr+>UBV
                        & APO  & UBVS & 16.81 & 17.25 & 17.15 &       &   6530\e24  & 3.12\e0.09 & $-$1.73\e0.30 & AV     & $-$176     &           \\ % SDSSJ015133.91+141105.2
SDSSJ015735.65+135254.2 & VLT  & SDSS & 17.92 & 17.80 & 19.01 & 0.061 &   8109\e156 & 3.10\e0.13 & $-$2.98\e0.15 & FHB    &    1\e2  & 30.4\e3.9 \\ % SDSSJ015735.65+135254.2  % OK, ugr+>UBV
                        &      & UBVS & 17.92 & 17.99 & 18.01 & \\
SDSSJ014631.99+133506.3 & VLT  & SDSS & 17.23 & 17.14 & 17.35 & 0.057 &             &            &             & QSO    &0.69$^{12}$&          \\ % SDSSJ014631.99+133506.3  % OK, ugr+>UBV
                        &      & UBVS & 17.26 & 17.32 & 16.60 & \\
2MASSJ195741.61$-$004009.7& VLT  & 2MS  & 16.56 & 16.90 &       & 0.240 &   6250\e250 & 4.50\e0.50 & $-$1.00\e0.25 & FV     &   73\e2  &  0.7\e0.6 \\ % 2MASSJ195741.61$-$004009.7 % GET RON 2MS
                        & APO  &      &       &       &       &       &   6250\e250 & 3.25\e0.25 & $-$1.25\e0.25 & FHB    &$-$$-$$^{13}$ & 11.1\e3.3 \\ % 2MASSJ195741.61$-$004009.7
2MASSJ195927.29+000822.3& VLT  & UBV  & 14.88 & 15.28 & 15.39 & 0.214 &   7755\e115 & 4.52\e0.25 & $-$1.01\e0.09 & AV     &   $-$1\e3  &  1.1\e0.5 \\ % 2MASSJ195927.29+000822.3 % OK, checked
                        & APO  &      &       &       &       &       &   7463\e257 & 4.41\e0.35 & $-$0.84\e0.26 & AV     &  $-$35     &  1.1\e0.5 \\ % 2MASSJ195927.29+000822.3
2MASSJ195925.49$-$000519.1& VLT  & UBV  & 14.24 & 14.70 & 14.93 & 0.215 &   7660\e45  & 4.18\e0.09 & $-$0.14\e0.07 & AV     &   34\e2  &  1.5\e0.2 \\ % 2MASSJ195925.49$-$000519.1 % OK, checked
                        & APO  &      &       &       &       &       &   7233\e124 & 3.67\e0.22 & $-$0.09\e0.08 & AV     &  $-$61     &  2.6\e0.4 \\ % 2MASSJ195925.49$-$000519.1
2MASSJ195912.00$-$002645.3& VLT  & UBV  & 13.96 & 14.42 & 14.52 & 0.156 &   7124\e92  & 3.73\e0.04 & $-$0.95\e0.06 & AV     &   24\e4  &  1.8\e0.3 \\ % 2MASSJ195912.00$-$002645.3 % OK, checked
                        & APO  &      &       &       &       &       &   7089\e88  & 4.03\e0.22 & $-$0.02\e0.06 & AV     &   27     &  1.7\e0.4 \\ % 2MASSJ195912.00$-$002645.3
2MASSJ195922.75+000300.0& VLT  & UBV  & 17.40 & 18.18 & 18.33 & 0.213 &   5762\e81  & 3.00\e0.50 &  0.00\e0.11 & FHB    &   21\e2  &  9.0\e3.7 \\ % 2MASSJ195922.75+000300.0 % OK, checked
                        & APO  &      &       &       &       &       &   5750\e250 & 3.00\e0.50 & $-$0.75\e0.25 & FHB/AV &   16     & 11.9\e3.9 \\ % 2MASSJ195922.75+000300.0
V1172Aql$^{14}$         & VLT  & Kuk  & 16.5  &       &       & 0.182 &   6500\e250 & 2.50\e0.50 & $-$0.85\e0.25 & RR     & $-$122\e1\v& 10.5\e1.4 \\ % V1172Aql                 % OK
2MASSJ195823.36$-$002719.0& VLT  & UBV  & 16.85 & 17.57 & 17.78 & 0.203 &   5911\e91  & 2.50\e0.50 & $-$0.35\e0.13 & FHB    &   40\e2  & 13.1\e3.2 \\ % 2MASSJ195823.36$-$002719.0 % OK, checked
                        & APO  &      &       &       &       &       &   6164\e42  & 2.39\e0.31 & $-$1.01\e0.13 & FHB    &   26     & 11.0\e3.0 \\ % 2MASSJ195823.36$-$002719.0
V1084Aql$^{15}$         & VLT  & Kuk  & 17.0  &       &       & 0.148 &   7000\e250 & 2.50\e0.50 & $-$1.22\e0.25 & RR     &  $-$32\e2  & 14.5\e1.3 \\ % V1084Aql                 % OK
BS17578$-$0015$^{16}$     & VLT  & GSC  & 10.19 & 10.45 &       & 0.110 &   7125\e250 & 2.50\e0.25 & $-$0.50\e0.25 & FHB    &   $-$9\e5  &  1.2\e0.3 \\ % BS17578$-$0015             % RON?
2MASSJ213520.04+133045.0& VLT  & UBV  & 13.22 & 13.44 & 13.74 & 0.131 &   7733\e58  & 4.00\e0.17 & $-$0.72\e0.15 & Am     &   26\e6  &  1.3\e0.2 \\ % 2MASSJ213520.04+133045.0 % OK, checked
                        & APO  &      &       &       &       &       &   7794\e180 & 4.00\e0.30 & $-$1.55\e0.08 & Am     &   $-$5\v   &  1.5\e0.5 \\ % 2MASSJ213520.04+133045.0
2MASSJ213451.58+134017.5& VLT  & UBV  & 13.62 & 13.90 & 14.16 & 0.106 &   7473\e171 & 4.00\e0.25 & $-$0.00\e0.02 & Am     &   22\e4  &  2.0\e0.6 \\ % 2MASSJ213451.58+134017.5 % OK, checked
                        & APO  &      &       &       &       &       &   7365\e17  & 4.00\e0.17 & $-$0.72\e0.15 & Am     &  $-$10     &  1.3\e0.3 \\ % 2MASSJ213451.58+134017.5
PG2134+125$^{17}$       & VLT  & UBV  & 17.02 & 17.45 & 17.08 & 0.131 &   $>$40000  &            &             & CSPN   &  $-$55\e9  &  2.2\e0.8 \\ % 2MASSJ213652.97+124719.0 % OK, checked
BS17578$-$0016$^{18}$     & VLT  & UBV  & 12.59 & 12.60 & 12.42 & 0.114 &  11772\e280 & 3.83\e0.10 &  0.00       & AV     &  $-$35\e5  &  3.6\e0.5 \\ % BS17578$-$0016             % OK, checked
                        & APO  &      &       &       &       &       &  12000\e200 & 4.00\e0.25 &  0.00\e0.25 & AV     &  $-$53     &  3.1\e0.8 \\ % BS17578$-$0016
FWPeg$^{19}$            & VLT  & Kuk  & 14.0  &       &       & 0.120 &   6500\e250 & 2.50\e0.50 & $-$0.82\e0.25 & RR     & $-$225\e2\v&  3.8\e0.8 \\ % FWPeg                    % OK
2MASSJ213327.04+133026.7& VLT  & UBV  & 14.98 & 15.02 & 14.90 & 0.111 &  11285\e176 & 3.97\e0.08 &  0.00       & AV     & $-$117\e3  &  8.2\e0.9 \\ % 2MASSJ213327.04+133026.7 % OK, checked
                        & APO  &      &       &       &       &       &  11500\e200 & 4.25\e0.25 &  0.00\e0.25 & AV     & $-$118     &  7.2\e0.4 \\ % 2MASSJ213327.04+133026.7
\enddata
\tablecomments{%
1: Source of spectroscopic data: APO=Apache Point Observatory; VLT=Very Large Telescope.
2: Source of photometry data: SDSS=Sloan Digital Sky Survey $ugr$; UBVS: SDSS photometry converted to UBV; UBV=our own UBV photometry; Str: Str\"omgren photometry; GSC: value from HST Guide Star Catalogue; Kuk: value from Kukarkin et al.\ (1970).
3: Magnitude -- $U$, $B$, $V$ when Col.~3 has ``UBV'', SDSS $r$, $g$, $u$ when Col.~3 has ``SDSS'', Str\"omgren $y$ when Col.~3 has ``Str'', $V$ magnitude from Kukarkin
et al.\ (1970) when Col.~3 has ``Kuk''.
4: Derived effective temperature.
5: Derived gravity.
6: Derived abundance, relative to solar.
7: Spectral type: FHB=Field Horizontal Branch; RR=RR\,Lyrae; AV, FV=main sequence A or F star.
8: Derived stellar velocity; a note (10) indicates that there are multiple spectra and the velocity varies; the listed velocity then refers to the velocity of the star
in the spectrum used to search for the interstellar absorption. Formal errors were not derived for velocities measured in the APO spectra -- the pixels are about 50~\kms\ wide.
9: Derived distance; we prefer the distance derived using the VLT spectrum; for RR\,Lyrae stars we estimate the average magnitude, as explained in the text.
10: Stellar velocity is variable, the listed velocity is valid for the spectrum in which the interstellar lines are searched-for.
11: Temperature, gravity and distance from Moehler et al.\ (1990); star located at R.A.,Dec.=01:45:39.5 +15:04:42.
12: Redshift instead of velocity given for this QSO.
13: Too many stellar lines to reliably determine stellar velocity in low-resolution spectrum.
14: Star located at R.A.,Dec.=20:02:23.0,$-$00:32:09.
15: Star located at R.A.,Dec.=20:03:43.5,+00:53:38.
16: Star located at R.A.,Dec.=21:34:17.8,+13:38:04
17: Distance from Phillips (2004); this is the central star of the planetary nebula NGC\,7094; star located at R.A.,Dec.=21:36:52.2,+13:35:00
18: Star located at R.A.,Dec.=21:34:45.4,+13:52:50
19: Star located at R.A.,Dec.=21:32:59.0,+12:52:24.
}
\end{deluxetable}

\def\e{$\pm$}
\def\kms{km\,s$^{-1}$}
\def\cmm#1{cm$^{-#1}$}
\def\CaII{\ion{Ca}{2}}
\def\NOTE#1{\hbox to 0pt{$^{#1}$\hss}}
\begin{deluxetable}{lrrrrcccc}
\tabletypesize{\footnotesize}
\tabcolsep=4pt \tablenum{2} \tablewidth{0pt}
\tablecolumns{9}
\tablecaption{Observational data}
\tablehead{%
\colhead{object} &\colhead{lon$^1$}  &\colhead{lat$^1$}  &\colhead{dist$^2$}&\colhead{T$_{\rm exp}^3$}&\colhead{Flux$^4$}&\colhead{S/N$^5$}&\colhead{Flux$^6$}&\colhead{S/N$^7$}    \\
                 &                   &                   &                  &                         &\colhead{(cont)}  &\colhead{cont}   &\colhead{(HVC)}   &\colhead{(@$v$(HVC))}\\
                 &\colhead{[$\circ$]}&\colhead{[$\circ$]}&\colhead{[kpc]}   &\colhead{[ks]}           &\colhead{[f.u.]}  &                 &\colhead{[f.u.]}  &                     \\
\colhead{(1)}    &\colhead{(2)}      &\colhead{(3)}      &\colhead{(4)}     &\colhead{(5)}            &\colhead{(6)}     &\colhead{(7)}    &\colhead{(8)}     &\colhead{(9)}         }
\startdata
\\\multicolumn4l{Cohen Stream}\\
PG0142+148                & 141.87 &$-$45.79 &     1.2\e0.5 &     0.5 &     226 &    84 &     226 &    78 \\
2MASSJ014825.88+132305.3  & 143.56 &$-$47.19 &     2.1\e1.3 &     1.0 &     137 &    83 &     128 &    80 \\
2MASSJ021651.34+080150.2  & 156.43 &$-$49.21 &     2.2\e1.0 & 26.1\NOTE{8} &     4.2 &    85 &     3.5 &    79 \\
2MASSJ021409.31+090105.3  & 154.81 &$-$48.67 &     3.3\e0.8 & 2.7\NOTE{9} &     190 &    84 &     132 &    55 \\
2MASSJ014936.48+143914.6  & 143.36 &$-$45.90 &     4.6\e0.8 &     1.3 &     114 &    80 &      54 &    38 \\
SDSSJ014843.61+130411.5   & 143.81 &$-$47.47 &     5.0\e1.2 &     3.0 &      45 &    96 &      15 &    42 \\
SDSSJ020033.50+141154.0   & 147.14 &$-$45.40 &     8.7\e1.3 & 8.0\NOTE{10} &      22 &    43 &      11 &    28 \\
SDSSJ015133.91+141105.2   & 144.23 &$-$46.19 &    11.2\e1.0 & 21.1\NOTE{11} &      17 &    99 &      13 &    69 \\
SDSSJ015735.65+135254.2   & 146.36 &$-$45.96 &    30.4\e3.9 & 15.0\NOTE{12} &     4.0 &    37 &     4.0 &    36 \\
SDSSJ014631.99+133506.3   & 142.83 &$-$47.15 &          *** &    14.0 &     7.9 &    51 &     7.9 &    51 \\
\\\multicolumn4l{Complex GCP}\\
2MASSJ195741.61-004009.7  &  40.04 &$-$15.01 &     0.7\e0.6 & 8.4\NOTE{13} &     5.6 &    37 &     0.8 &   7.6 \\
2MASSJ195927.29+000822.3  &  41.00 &$-$15.02 &     1.1\e0.5 &     1.6 &      38 &    52 &      17 &    35 \\
2MASSJ195925.49-000519.1  &  40.79 &$-$15.12 &     1.5\e0.2 &     1.0 &      67 &    42 &      10 &    14 \\
2MASSJ195912.00-002645.3  &  40.43 &$-$15.24 &     1.8\e0.3 &     1.0 &      84 &    55 &      22 &    25 \\
2MASSJ195922.75+000300.0  &  40.91 &$-$15.05 &     9.0\e3.7 &     1.7 &     2.5 &    11 &     0.5 &   3.3 \\
V1172Aql                  &  40.74 &$-$15.98 &    10.5\e1.4 & 8.2\NOTE{14} &      11 &    52 &     7.6 &    52 \\
2MASSJ195823.36-002719.0  &  40.32 &$-$15.07 &    13.1\e3.2 & 5.6\NOTE{15} &     3.9 &    24 &     0.5 &   4.6 \\
V1084Aql                  &  42.21 &$-$15.60 &    14.5\e1.3 & 11.0\NOTE{16} &      10 &    55 &     8.6 &    47 \\
\\\multicolumn4l{Cloud g1}\\
BS17578-0015              &  67.04 &$-$27.17 &     1.2\e0.3 &     0.5 &     2.5 &    77 &     1.9 &    70 \\
2MASSJ213520.04+133045.0  &  67.12 &$-$27.44 &     1.3\e0.2 &     1.5 &     160 &    81 &      59 &    34 \\
2MASSJ213451.58+134017.5  &  67.17 &$-$27.25 &     2.0\e0.6 & 2.6\NOTE{17} &      84 &   101 &      41 &    47 \\
PG2134+125                &  66.78 &$-$28.20 &     2.2\e0.8 &     3.0 &     249 &   103 &     241 &   148 \\
BS17578-0016              &  67.33 &$-$27.10 &     3.6\e0.5 &     0.5 &     658 &   106 &     603 &   101 \\
FWPeg                     &  66.15 &$-$27.39 &     3.8\e0.8 & 3.0\NOTE{18} &      67 &    71 &      64 &    73 \\
2MASSJ213327.04+133026.7  &  66.78 &$-$27.10 &     8.2\e0.9 & 3.0\NOTE{19} &      40 &    73 &      40 &    91 \\
\enddata
\tablecomments{%
1: Galactic longitude and latitude of the star.
2: Distance estimate from Table~1.
3: Total exposure time, in kiloseconds (but see notes 8 to 19).
4: Stellar flux in the continuum in units of 10$^{-16}$\,erg\,\cmm2\,s$^{-1}$\,\AA$^{-1}$.
5: Signal-to-noise ratio in the continuum.
6: Stellar flux at the velocity of the high-velocity cloud in units of 10$^{-16}$~erg\,\cmm2\,s$^{-1}$\,\AA$^{-1}$.
7: Signal-to-noise ratio at the velocity of the high-velocity cloud.
%
% PG0142+148
8: There are nine exposures -- since the stellar velocity is constant they are all combined. % 2MASSJ021651.34+080150.2
9: There are four exposures of this star, but for unknown reasons the flux in the second is much lower, so we exclude it. % 2MASSJ021409.31+090105.3
% 2MASSJ014825.88+132305.3
% 2MASSJ014936.48+143914.6
% SDSSJ014843.61+130411.5
10: There are three exposures; the stellar velocity is $-$2~\kms\ in the first exposure, $-$8~\kms\ in the second, $-$29~\kms\ in the third -- we only use the third exposure, as
in exposures 1 and 2 a stellar line is shifted on top of the interstellar \CaII~K line. % SDSSJ020033.50+141154.0
11: This star appears to be binary: in four of the seven exposures the stellar lines are at $-$190~\kms\ (and we combine these four), but in the remaining three exposures the
stellar line is much broader, appears to have two components, and is centered at $-$178, $-$138 and $-$139~\kms; it also confuses the interstellar \CaII~K. % SDSSJ015133.91+141105.2
12: There are five exposures, with constant stellar velocity; however in the first exposure a spike (probably a cosmic ray) confuses the interstellar K line. % SDSSJ015735.65+135254.2
% SDSSJ014631.99+133506.3
13: The first of the three exposures has lower stellar flux, so we don't use it. % 2MASSJ195741.61-004009.7
% 2MASSJ195927.29+000822.3
% 2MASSJ195925.49-000519.1
% 2MASSJ195912.00-002645.3
% 2MASSJ195922.75+000300.3
14: The stellar velocity changed from $-$128 to $-$121 to $-$115~\kms\ over the three contiguous hours of observing; since there are no stellar lines interfering with the
interstellar \CaII~K, we averaged the three exposures. % V1172Aql
15: The two exposures of this star were combined. % 2MASSJ195823.36-002719.0
16: The stellar velocity changed from $-$36 to $-$31 to $-$26~\kms\ over the four contiguous hours of observing; since there are no stellar lines interfering with the
interstellar \CaII~K, we averaged these. % V1084Aql
% BS17578-0015
% 2MASSJ213520.04+133045.0
17: The two exposures of this star were combined. % 2MASSJ213451.58+134017.5
% PG2134+125=2MASSJ213652.97+124719.0
% BS17578-0016
18: The stellar velocity is $-$225~\kms\ in the first exposure, $-$170~\kms\ in the second; the wide stellar \CaII~K line extends across the interstellar line in the second
exposure, halving the flux, so we only use the first exposure. % FWPeg
19: The two exposures of this star were combined. % 2MASSJ213327.04+133026.7
}
\end{deluxetable}

\def\e{$\pm$}
\def\kms{km\,s$^{-1}$}
\def\cmm#1{cm$^{-#1}$}
\def\HI{\ion{H}{1}}
\def\CaII{\ion{Ca}{2}}
\def\FeI{\ion{Fe}{1}}
\def\Wexp{W$_{exp}^3$}
\def\Wobs{W$_{obs}^3$}
\def\NOTE#1{\hbox to 0pt{$^{#1}$\hss}}
\begin{deluxetable}{lrccrrrrrrrrc}
\rotate
\tabletypesize{\footnotesize}
\tabcolsep=2pt \tablenum{3} \tablewidth{0pt}
\tablecolumns{13}
\tablecaption{Interstellar absorption line results}
\tablehead{%
\colhead{object} &\colhead{dist}    &\colhead{$v_{\rm LSR}^1$}&\colhead{log$^2$}    &\colhead{\Wexp}      &\colhead{\Wexp}  &\colhead{\Wobs} &\colhead{\Wobs}
                 &\colhead{log$N^4$}&\colhead{log$N^4$}       &\colhead{$A^5$}      &\colhead{$A^5$}      &\colhead{U,L$^6$}\\
                 &\colhead{[kpc]}   &\colhead{[\kms]}         &\colhead{$N$(HI)}    &\colhead{[m\AA]}     &\colhead{[m\AA]} &\colhead{[m\AA]}&\colhead{[m\AA]}
                 &                  &                         &\colhead{[10$^{-9}$]}&\colhead{[10$^{-9}$]}&\colhead{}       \\
                 &                  &                         &                     &\colhead{(K)}        &\colhead{(H)}    &\colhead{(K)}   &\colhead{(H)}
                 &\colhead{(K)}     &\colhead{(H)}            &\colhead{(K)}        &\colhead{(H)}        &\colhead{}       \\
\colhead{(1)}    &\colhead{(2)}     &\colhead{(3)}            &\colhead{(4)}        &\colhead{(5)}        &\colhead{(6)}    &\colhead{(7)}   &\colhead{(8)}
                 &\colhead{(9)}     &\colhead{(10)}           &\colhead{(11)}       &\colhead{(12)}       &\colhead{(13)}   }
\startdata
\multicolumn{10}l{Cohen Stream (D=5.0--11.7~kpc)}\\
PG0142+148                &     1.2\e0.5 &$-$105 &  19.26 & 20 &  10    &                 $<$7 &                 $<$8 &             $<$10.88 &             $<$11.38 &               $<$4.2 &              $<$13.2 & L  \\%              $<$0.62               $<$1.12                 26.7
2MASSJ014825.88+132305.3  &     2.1\e1.3 &$-$106 &  19.41 & 22 &  11    &                 $<$6 &                $<$15 &             $<$10.92 &             $<$11.38 &               $<$3.2 &               $<$9.3 & L  \\%              $<$0.51               $<$0.97                 20.4
2MASSJ021651.34+080150.2  &     2.2\e1.0 &$-$112 &  19.54 & 24 &  12    &                $<$10 &                 $<$9 &             $<$11.05 &             $<$11.37 &               $<$3.2 &               $<$6.8 & L  \\%              $<$0.51               $<$0.83                 16.2
2MASSJ021409.31+090105.3  &     3.3\e0.8 &$-$110 &  19.08 & 19 &   9    &                  \ &                  \ &                  \ &                  \ &                  \ &                  \ & \  \\%                  \                   \                 36.9
2MASSJ014936.48+143914.6  &     4.6\e0.8 &$-$106 &  19.30 & 21 &  10    &                 $<$9 &                $<$11 &             $<$11.02 &             $<$11.48 &               $<$5.2 &                $<$15 & L  \\%              $<$0.72               $<$1.18                 24.9
SDSSJ014843.61+130411.5   &     5.0\e1.2 &$-$103 &  19.18 & 19 &  10    &                $<$10 &                $<$10 &             $<$11.08 &             $<$11.48 &               $<$7.9 &                $<$20 & \  \\%              $<$0.90               $<$1.30                 30.9
SDSSJ020033.50+141154.0   &     8.7\e1.3 &$-$108 &  18.94 & 17 &   9    &                $<$17 &                $<$19 &             $<$11.28 &             $<$11.72 &                $<$22 &                $<$60 & \  \\%              $<$1.34               $<$1.78                 47.5
SDSSJ015133.91+141105.2   &    11.2\e1.0 &$-$104 &  19.26 & 20 &  10    &           17$\pm$3 &                $<$28 &     11.33$\pm$0.07 &             $<$11.54 &       11.7$\pm$2.9 &                $<$19 & U  \\%               1.07               $<$1.28                 26.7
SDSSJ015735.65+135254.2   &    30.4\e3.9 &$-$104 &  18.90 & 17 &   8    &           10$\pm$3 &                $<$18 &     11.06$\pm$0.20 &             $<$11.63 &       14.5$\pm$7.6 &                $<$54 & U  \\%               1.16               $<$1.73                 51.1
SDSSJ014631.99+133506.3   &          *** &$-$105 &  19.51 & 23 &  11    &           17$\pm$3 &           11$\pm$3 &     11.31$\pm$0.09 &     11.41$\pm$0.12 &        6.3$\pm$1.7 &        7.9$\pm$2.5 & U  \\%               0.80                0.90                 17.1
\\\multicolumn{10}l{IV-South (D=1.0--2.7~kpc)}\\
PG0142+148                &     1.2\e0.5 &$-$51 &  19.54 & 28 &  14    &                 $<$7 &                 $<$8 &             $<$10.94 &             $<$11.25 &               $<$2.5 &               $<$5.1 & L  \\%              $<$0.40               $<$0.71                 16.2
2MASSJ014825.88+132305.3  &     2.1\e1.3 &$-$41 &  19.11 & 23 &  11    &           25$\pm$2 &                  \ &     11.50$\pm$0.04 &                  \ &           24$\pm$5 &                  \ & U  \\%               1.39                   \                 35.0
2MASSJ021651.34+080150.2  &     2.2\e1.0 &$-$75 &  18.93 & 21 &  10    &                  \ &                  \ &                  \ &                  \ &                  \ &                  \ & \  \\%                  \                   \                 48.4
2MASSJ021409.31+090105.3  &     3.3\e0.8 &$-$63 &  18.77 & 19 &   9    &                  \ &                  \ &                  \ &                  \ &                  \ &                  \ & \  \\%                  \                   \                 64.5
2MASSJ014936.48+143914.6  &     4.6\e0.8 &$-$51 &  19.49 & 27 &  14    &           26$\pm$4 &                $<$14 &     11.50$\pm$0.06 &             $<$11.50 &       10.2$\pm$2.5 &              $<$10.2 & U  \\%               1.01               $<$1.01                 17.7
SDSSJ014843.61+130411.5   &     5.0\e1.2 &$-$35 &  19.18 & 23 &  12    &                $<$12 &                $<$10 &             $<$11.02 &             $<$11.49 &               $<$6.9 &                $<$20 & \  \\%              $<$0.84               $<$1.31                 30.9
SDSSJ020033.50+141154.0   &     8.7\e1.3 &$-$46 &  18.86 & 20 &  10    &                  \ &                  \ &                  \ &                  \ &                  \ &                  \ & \  \\%                  \                   \                 54.9
SDSSJ015133.91+141105.2   &    11.2\e1.0 &$-$49 &  19.23 & 24 &  12    &           16$\pm$3 &           26$\pm$2 &     11.29$\pm$0.08 &     11.50$\pm$0.04 &       11.5$\pm$3.0 &           19$\pm$4 & U  \\%               1.06                1.27                 28.2
SDSSJ015735.65+135254.2   &    30.4\e3.9 &$-$48 &  18.95 & 21 &  10    &           22$\pm$3 &                $<$21 &     11.42$\pm$0.10 &             $<$11.80 &           30$\pm$9 &                $<$71 & U  \\%               1.47               $<$1.85                 46.7
SDSSJ014631.99+133506.3   &          *** &$-$43 &  19.20 & 24 &  12    &           30$\pm$3 &           19$\pm$3 &     11.55$\pm$0.06 &     11.66$\pm$0.10 &           22$\pm$5 &           29$\pm$8 & U  \\%               1.35                1.46                 29.8
\\\multicolumn{10}l{Complex GCP (D=9.8-15.1~kpc)}\\
2MASSJ195741.61-004009.7  &     0.7\e0.6 & 89 &  20.16 & 61 &  32    &                $<$43 &                $<$40 &             $<$11.70 &             $<$12.19 &               $<$3.5 &              $<$10.7 & \  \\%              $<$0.54               $<$1.03                  5.3
2MASSJ195927.29+000822.3  &     1.1\e0.5 & 93 &  20.19 & 64 &  33    &                $<$13 &                $<$18 &             $<$11.17 &             $<$11.61 &               $<$1.0 &               $<$2.6 & L  \\%             $<$$-$0.02               $<$0.42                  5.0
2MASSJ195925.49-000519.1  &     1.5\e0.2 & 93 &  20.15 & 62 &  32    &                $<$28 &                  \ &             $<$11.58 &                  \ &               $<$2.7 &                  \ & \  \\%              $<$0.43                   \                  5.4
2MASSJ195912.00-002645.3  &     1.8\e0.3 & 93 &  20.13 & 62 &  32    &                $<$15 &                $<$16 &             $<$11.18 &             $<$11.49 &               $<$1.1 &               $<$2.3 & L  \\%              $<$0.05               $<$0.36                  5.6
2MASSJ195922.75+000300.0  &     9.0\e3.7 & 93 &  20.17 & 63 &  32    &               $<$124 &               $<$158 &             $<$12.36 &             $<$12.54 &                $<$16 &                $<$23 & \  \\%              $<$1.19               $<$1.37                  5.2
V1172Aql                  &    10.5\e1.4 & 94 &  20.03 & 59 &  30    &                $<$11 &                $<$13 &             $<$11.13 &             $<$11.56 &               $<$1.3 &               $<$3.4 & L  \\%              $<$0.10               $<$0.53                  6.7
2MASSJ195823.36-002719.0  &    13.1\e3.2 & 90 &  20.12 & 61 &  31    &                $<$68 &                $<$57 &             $<$12.09 &             $<$12.25 &               $<$9.3 &              $<$13.5 & \  \\%              $<$0.97               $<$1.13                  5.7
V1084Aql                  &    14.5\e1.3 & 91 &  20.08 & 61 &  31    &           51$\pm$3 &          37$\pm$10 &     11.82$\pm$0.02 &     11.96$\pm$0.13 &        5.5$\pm$1.1 &        7.6$\pm$2.4 & U  \\%               0.74                0.88                  6.1
\\\multicolumn{10}l{Cloud g1 (D=1.8-3.8~kpc)}\\
BS17578-0015              &     1.2\e0.3 & 71 &  19.04 & 74 &  39    &                 $<$5 &                 $<$8 &             $<$10.73 &             $<$11.02 &               $<$4.9 &               $<$9.5 & L  \\%              $<$0.69               $<$0.98                 39.7
2MASSJ213520.04+133045.0  &     1.3\e0.2 & 71 &  19.04 & 75 &  39    &                 $<$8 &                $<$11 &             $<$10.99 &             $<$11.59 &               $<$8.9 &                $<$36 & L  \\%              $<$0.95               $<$1.55                 39.7
2MASSJ213451.58+134017.5  &     2.0\e0.6 & 71 &  18.95 & 72 &  38    &                 $<$8 &                $<$10 &             $<$10.94 &             $<$11.35 &               $<$9.8 &                $<$25 & L  \\%              $<$0.99               $<$1.40                 46.7
PG2134+125                &     2.2\e0.8 & 70 &  18.92 & 69 &  37    &                 $<$4 &                 $<$5 &             $<$10.66 &             $<$11.00 &               $<$5.5 &              $<$12.0 & L  \\%              $<$0.74               $<$1.08                 49.2
BS17578-0016              &     3.6\e0.5 & 67 &  18.51 & 55 &  29    &           30$\pm$2 &           14$\pm$3 &     11.56$\pm$0.03 &     11.52$\pm$0.14 &         112$\pm$24 &         102$\pm$35 & U  \\%               2.05                2.01                102.8
\                         &            \ & 80 &  18.08 & 42 &  23    &           44$\pm$2 &           31$\pm$3 &     11.74$\pm$0.02 &     11.87$\pm$0.04 &         457$\pm$94 &        617$\pm$137 & U  \\%               2.66                2.79                222.6
FWPeg                     &     3.8\e0.8 & 67 &  19.23 & 82 &  43    &          127$\pm$4 &           73$\pm$4 &     12.25$\pm$0.01 &     12.26$\pm$0.03 &         105$\pm$21 &         107$\pm$23 & U  \\%               2.02                2.03                 28.2
2MASSJ213327.04+133026.7  &     8.2\e0.9 & 72 &  19.11 & 77 &  41    &           66$\pm$3 &           27$\pm$9 &     11.92$\pm$0.02 &     11.88$\pm$0.14 &          65$\pm$13 &          59$\pm$20 & U  \\%               1.81                1.77                 35.0
\\\multicolumn{10}l{Low-velocity gas near CS }\\
PG0142+148                &     1.2\e0.5 &$-$3 &  20.56 & 203 &  112    &           94$\pm$2 &           65$\pm$4 &     12.12$\pm$0.01 &     12.22$\pm$0.03 &        3.6$\pm$0.7 &        4.6$\pm$1.0 & U  \\%               0.56                0.66                  2.6
2MASSJ014825.88+132305.3  &     2.1\e1.3 &$-$6 &  20.65 & 208 &  115    &           98$\pm$3 &           64$\pm$7 &     12.16$\pm$0.01 &     12.23$\pm$0.05 &        3.2$\pm$0.6 &        3.8$\pm$0.9 & U  \\%               0.51                0.58                  2.2
2MASSJ021651.34+080150.2  &     2.2\e1.0 &$-$12 &  20.81 & 226 &  127    &          115$\pm$6 &                $<$11 &     12.24$\pm$0.02 &             $<$12.23 &        2.7$\pm$0.6 &               $<$2.6 & U  \\%               0.43               $<$0.42                  1.7
2MASSJ021409.31+090105.3  &     3.3\e0.8 &$-$13 &  20.75 & 213 &  119    &                  \ &                  \ &                  \ &                  \ &                  \ &                  \ & \  \\%                  \                   \                  1.8
2MASSJ014936.48+143914.6  &     4.6\e0.8 &$-$4 &  20.59 & 206 &  114    &          124$\pm$5 &           52$\pm$7 &     12.26$\pm$0.02 &     12.13$\pm$0.07 &        4.7$\pm$1.0 &        3.5$\pm$0.9 & U  \\%               0.67                0.54                  2.5
SDSSJ014843.61+130411.5   &     5.0\e1.2 &$-$5 &  20.69 & 216 &  121    &          120$\pm$5 &           78$\pm$4 &     12.26$\pm$0.02 &     12.31$\pm$0.04 &        3.7$\pm$0.8 &        4.2$\pm$0.9 & U  \\%               0.57                0.62                  2.1
SDSSJ020033.50+141154.0   &     8.7\e1.3 &$-$4 &  20.68 & 212 &  118    &         120$\pm$18 &         102$\pm$16 &     12.30$\pm$0.08 &     12.49$\pm$0.06 &        4.2$\pm$1.1 &        6.5$\pm$1.6 & U  \\%               0.62                0.81                  2.1
SDSSJ015133.91+141105.2   &    11.2\e1.0 &$-$6 &  20.62 & 204 &  113    &          183$\pm$3 &          123$\pm$7 &     12.57$\pm$0.01 &     12.61$\pm$0.03 &        8.9$\pm$1.8 &        9.8$\pm$2.0 & U  \\%               0.95                0.99                  2.3
SDSSJ015735.65+135254.2   &    30.4\e3.9 &$-$6 &  20.66 & 213 &  118    &                  \ &                  \ &                  \ &                  \ &                  \ &                  \ & \  \\%                  \                   \                  2.2
SDSSJ014631.99+133506.3   &          *** &$-$4 &  20.47 & 193 &  106    &          155$\pm$4 &           76$\pm$4 &     12.35$\pm$0.02 &     12.30$\pm$0.03 &        7.6$\pm$1.5 &        6.8$\pm$1.5 & U  \\%               0.88                0.83                  3.0
\\\multicolumn{10}l{Low-velocity gas near GCP }\\
2MASSJ195741.61-004009.7  &     0.7\e0.6 & 32 &  20.47 & 177 &  101    &         181$\pm$16 &                $<$74 &     12.50$\pm$0.06 &             $<$12.35 &       10.7$\pm$2.5 &               $<$7.6 & U  \\%               1.03               $<$0.88                  3.0
2MASSJ195927.29+000822.3  &     1.1\e0.5 & 30 &  20.43 & 176 &  101    &          186$\pm$6 &           89$\pm$9 &     12.52$\pm$0.02 &     12.41$\pm$0.05 &       12.3$\pm$2.5 &        9.5$\pm$2.1 & U  \\%               1.09                0.98                  3.3
2MASSJ195925.49-000519.1  &     1.5\e0.2 & 32 &  20.40 & 176 &  101    &                  \ &                  \ &                  \ &                  \ &                  \ &                  \ & \  \\%                  \                   \                  3.5
2MASSJ195912.00-002645.3  &     1.8\e0.3 & 35 &  20.37 & 170 &  97    &         124$\pm$10 &           84$\pm$9 &     12.24$\pm$0.05 &     12.35$\pm$0.05 &        7.4$\pm$1.7 &        9.5$\pm$2.2 & U  \\%               0.87                0.98                  3.6
2MASSJ195922.75+000300.0  &     9.0\e3.7 & 32 &  20.42 & 174 &  99    &         366$\pm$46 &               $<$222 &     12.91$\pm$0.15 &             $<$12.84 &          31$\pm$16 &                $<$26 & U  \\%               1.49               $<$1.42                  3.3
V1172Aql                  &    10.5\e1.4 & 27 &  20.41 & 175 &  100    &          230$\pm$4 &          152$\pm$6 &     12.58$\pm$0.01 &     12.67$\pm$0.02 &       14.8$\pm$3.0 &           18$\pm$4 & U  \\%               1.17                1.26                  3.4
2MASSJ195823.36-002719.0  &    13.1\e3.2 & 32 &  20.40 & 175 &  100    &         303$\pm$24 &         250$\pm$22 &     12.87$\pm$0.13 &     12.96$\pm$0.07 &          30$\pm$12 &           36$\pm$9 & U  \\%               1.47                1.56                  3.5
V1084Aql                  &    14.5\e1.3 & 32 &  20.33 & 167 &  95    &          240$\pm$3 &         130$\pm$11 &     12.65$\pm$0.01 &     12.57$\pm$0.04 &           21$\pm$4 &           17$\pm$4 & U  \\%               1.32                1.24                  3.9
\\\multicolumn{10}l{Low-velocity gas near GCP }\\
2MASSJ195741.61-004009.7  &     0.7\e0.6 & 0 &  20.82 & 200 &  117    &         118$\pm$11 &          94$\pm$13 &     12.31$\pm$0.06 &     12.46$\pm$0.06 &        3.1$\pm$0.7 &        4.4$\pm$1.1 & U  \\%               0.49                0.64                  1.6
2MASSJ195927.29+000822.3  &     1.1\e0.5 & 0 &  20.85 & 202 &  118    &          213$\pm$5 &          121$\pm$7 &     12.61$\pm$0.02 &     12.56$\pm$0.03 &        5.8$\pm$1.2 &        5.1$\pm$1.1 & U  \\%               0.76                0.71                  1.5
2MASSJ195925.49-000519.1  &     1.5\e0.2 & 0 &  20.82 & 202 &  118    &                  \ &                  \ &                  \ &                  \ &                  \ &                  \ & \  \\%                  \                   \                  1.6
2MASSJ195912.00-002645.3  &     1.8\e0.3 & 0 &  20.79 & 199 &  116    &          154$\pm$9 &          111$\pm$8 &     12.42$\pm$0.01 &     12.50$\pm$0.03 &        4.3$\pm$0.9 &        5.1$\pm$1.1 & U  \\%               0.63                0.71                  1.7
2MASSJ195922.75+000300.0  &     9.0\e3.7 & 0 &  20.84 & 207 &  122    &         193$\pm$43 &               $<$160 &     12.60$\pm$0.31 &             $<$12.71 &        5.8$\pm$3.2 &               $<$7.4 & U  \\%               0.76               $<$0.87                  1.6
V1172Aql                  &    10.5\e1.4 & 1 &  20.78 & 196 &  114    &          128$\pm$5 &          182$\pm$5 &     12.32$\pm$0.02 &     12.78$\pm$0.02 &        3.5$\pm$0.7 &       10.0$\pm$2.0 & U  \\%               0.54                1.00                  1.7
2MASSJ195823.36-002719.0  &    13.1\e3.2 & 0 &  20.80 & 203 &  119    &         200$\pm$18 &         150$\pm$17 &     12.67$\pm$0.16 &     12.73$\pm$0.07 &        7.4$\pm$3.5 &        8.5$\pm$2.2 & U  \\%               0.87                0.93                  1.7
V1084Aql                  &    14.5\e1.3 & 1 &  20.82 & 199 &  117    &          248$\pm$4 &          181$\pm$9 &     12.67$\pm$0.01 &     12.74$\pm$0.02 &        7.1$\pm$1.4 &        8.3$\pm$1.7 & U  \\%               0.85                0.92                  1.6
\\\multicolumn{10}l{Low-velocity gas near g1 }\\
BS17578-0015              &     1.2\e0.3 & 2 &  20.82 & 221 &  123    &          115$\pm$3 &           74$\pm$3 &     12.25$\pm$0.01 &     12.30$\pm$0.02 &        2.7$\pm$0.5 &        3.0$\pm$0.6 & U  \\%               0.43                0.48                  1.6
2MASSJ213520.04+133045.0  &     1.3\e0.2 & 4 &  20.84 & 228 &  128    &          235$\pm$3 &          176$\pm$3 &     12.66$\pm$0.01 &     12.74$\pm$0.01 &        6.6$\pm$1.3 &        7.9$\pm$1.6 & U  \\%               0.82                0.90                  1.6
2MASSJ213451.58+134017.5  &     2.0\e0.6 & 4 &  20.82 & 220 &  123    &          155$\pm$9 &          108$\pm$6 &     12.39$\pm$0.02 &     12.48$\pm$0.02 &        3.7$\pm$0.8 &        4.6$\pm$0.9 & U  \\%               0.57                0.66                  1.6
PG2134+125                &     2.2\e0.8 & 1 &  20.84 & 230 &  129    &          178$\pm$2 &          107$\pm$2 &     12.46$\pm$0.00 &     12.47$\pm$0.01 &        4.2$\pm$0.8 &        4.3$\pm$0.9 & U  \\%               0.62                0.63                  1.6
BS17578-0016              &     3.6\e0.5 & 2 &  20.79 & 219 &  123    &          216$\pm$2 &          132$\pm$3 &     12.58$\pm$0.00 &     12.58$\pm$0.01 &        6.2$\pm$1.2 &        6.2$\pm$1.3 & U  \\%               0.79                0.79                  1.7
FWPeg                     &     3.8\e0.8 & 3 &  20.79 & 197 &  116    &          287$\pm$3 &          197$\pm$5 &     12.72$\pm$0.01 &     12.78$\pm$0.01 &        8.5$\pm$1.7 &        9.8$\pm$2.0 & U  \\%               0.93                0.99                  1.7
2MASSJ213327.04+133026.7  &     8.2\e0.9 & 2 &  20.80 & 226 &  127    &          260$\pm$2 &                  \ &     12.65$\pm$0.00 &                  \ &        7.1$\pm$1.4 &                  \ & U  \\%               0.85                   \                  1.7
\enddata
\tablecomments{%
1: Velocity of HVC or IVC in the direction of the star.
2: Logarithm of 21-cm \HI\ column density in the high-velocity cloud.
3: Expected and observed equivalent width of the high-velocity cloud \CaII\ absorption.
4: Logarithm of integrated apparent \CaII\ column density (in \cmm2) based on the K (Col.~9) or H (Col.~10) line profile.
5: \CaII/\HI\ ratio in units of 10$^{-9}$, using $N$(K) (Col.~11) or $N$(H) (Col.~12); the error includes a 20\% error in the \HI\ column density.
6: Conclusion drawn concerning HVC or IVC distance from this sightline: U means upper limit, L means lower limit, no entry is given if the non-detection is not significant.
}
\end{deluxetable}

\def\X{$\times$}
\def\tdex#1{$\times$10$^{#1}$}
\def\cmm#1{cm$^{-#1}$}
\def\HI{H\,I}
\def\Mdot{M}
\def\Msun{M$_\odot$}
\def\Msunpyr{M$_\odot$\,yr$^{-1}$}
\begin{deluxetable}{lcccc}
\tabletypesize{\footnotesize}
\tabcolsep=5pt \tablenum{4} \tablewidth{0pt}
\tablecolumns{5}
\tablecaption{Summary of cloud parameters}
\tablehead{%
\colhead{Quantity}      &\colhead{Cohen Stream}    &\colhead{Perseus Arm IVC}   &\colhead{Complex~GCP}  &\colhead{Cloud~g1}           }
\startdata
Distance [kpc]          &        5.0--11.7        &        1.0--2.7         &       9.8--15.1         &        1.8--3.8         \\
Radius [kpc]            &       11.5--15.7        &        9.1--10.2        &       6.2--9.7          &        8.0--7.8         \\
Height [kpc]            &        3.7--8.6         &        0.7--1.8         &       2.5--3.9          &        0.8--1.7         \\
Size [deg]              &          1\X12          &          3\X8           &         2\X8            &          2\X2           \\
\\
\HI\ parameters \\
$N$(HI) (peak) [\cmm2]  &           3.5\tdex{19}  &           6.0\tdex{19}  &          17.0\tdex{19}  &           2.5\tdex{19}  \\
$Lx$ [pc]               &         90--200         &         50--140         &        340--530         &         60--130         \\
$Ly$ [pc]               &       1000--2500        &        100--400         &       1400--2100        &         60--130         \\
$n$(HI) [\cmm3]         &       0.13--0.06        &       0.37--0.14        &       0.16--0.10        &       0.13--0.06        \\
$M$(HI) [\Msun]         &      0.73--4.02\tdex{ 5} &     1.86--13.57\tdex{ 4} &      0.68--1.60\tdex{ 6} &      0.09--0.41\tdex{ 4} \\
\Mdot(HI) [\Msunpyr]    &       2.0--4.7\tdex{-3} &     16.0--43.2\tdex{-4} &       6.4--9.8\tdex{-3} &       0.8--1.6\tdex{-4} \\
\\
\multicolumn5l{H$^+$ parameters assuming H$^+$ has same pathlength as \HI}\\
$I$(H$\alpha$) [R]      &       0.07$\pm$0.01     &       $<$0.15     &       0.27$\pm$0.03       &       1.1$\pm$0.2       \\
$n$(H$^+$) [\cmm3]      &       0.04--0.03        &      $<$0.08--$<$0.05      &       0.04--0.03        &       0.20--0.14        \\
$N$(H$^+$) [\cmm2]      &       1.1--1.7\tdex{19} &    $<$1.3--$<$2.1\tdex{19} &       4.4--5.5\tdex{19} &       3.8--5.6\tdex{19} \\
$M$(H$^+$) [\Msun]      &       0.7--5.9\tdex{ 5} &   $<$0.7--$<$8.5\tdex{ 4} &       0.6--1.9\tdex{ 6} &       0.4--2.3\tdex{ 4} \\
\Mdot(H$^+$) [\Msunpyr] &       1.9--6.8\tdex{-3} &   $<$6.0--$<$26.8\tdex{-4} &       6.7--12.8\tdex{-3} &       3.0--9.2\tdex{-4} \\
\\
\multicolumn5l{H$^+$ parameters assuming H$^+$ has same volume density as \HI}\\
$L$ [pc]                &          9--51          &       $<$2--$<$18       &         23--55          &        148--660         \\
$n$(H$^+$) [\cmm3]      &       0.13--0.06        &      $<$0.37--$<$0.14      &       0.16--0.10        &       0.13--0.06        \\
$N$(H$^+$) [\cmm2]      &       0.4--0.9\tdex{19} &    $<$0.3--$<$0.8\tdex{19} &       1.2--1.8\tdex{19} &      5.9--12.4\tdex{19} \\
$M$(H$^+$) [\Msun]      &       0.2--2.9\tdex{ 5} &   $<$0.2--$<$3.0\tdex{ 4} &       0.2--0.6\tdex{ 6} &       0.6--5.2\tdex{ 4} \\
\Mdot(H$^+$) [\Msunpyr] &       0.6--3.4\tdex{-3} &   $<$1.3--$<$9.5\tdex{-4} &       1.7--4.1\tdex{-3} &       4.6--20.5\tdex{-4} \\
\enddata
\tablecomments{%
The mass flows are for infall in the case of the Cohen Stream and the Perseus
Arm IVC, but they represent outflows for complex~GCP and cloud~g1.
}
\end{deluxetable}

\clearpage

%%%%%%%%%%%%%%%%%%%%%%%%%%%%%%%%%%%%%%%%%%%%%%%%%%%%%%%%%%%%%%%%%%%%%%%%%%%%%%%%
%%%%%%%%%%%%%%%%%%%%%%%%%%%%%%%%%%%%%%%%%%%%%%%%%%%%%%%%%%%%%%%%%%%%%%%%%%%%%%%%

\clearpage

%plotfiddle(1=file 2=box 3=angle 4=hscale 5=vscale 6=hoffset 7=voffset)
\begin{figure}\plotfiddle{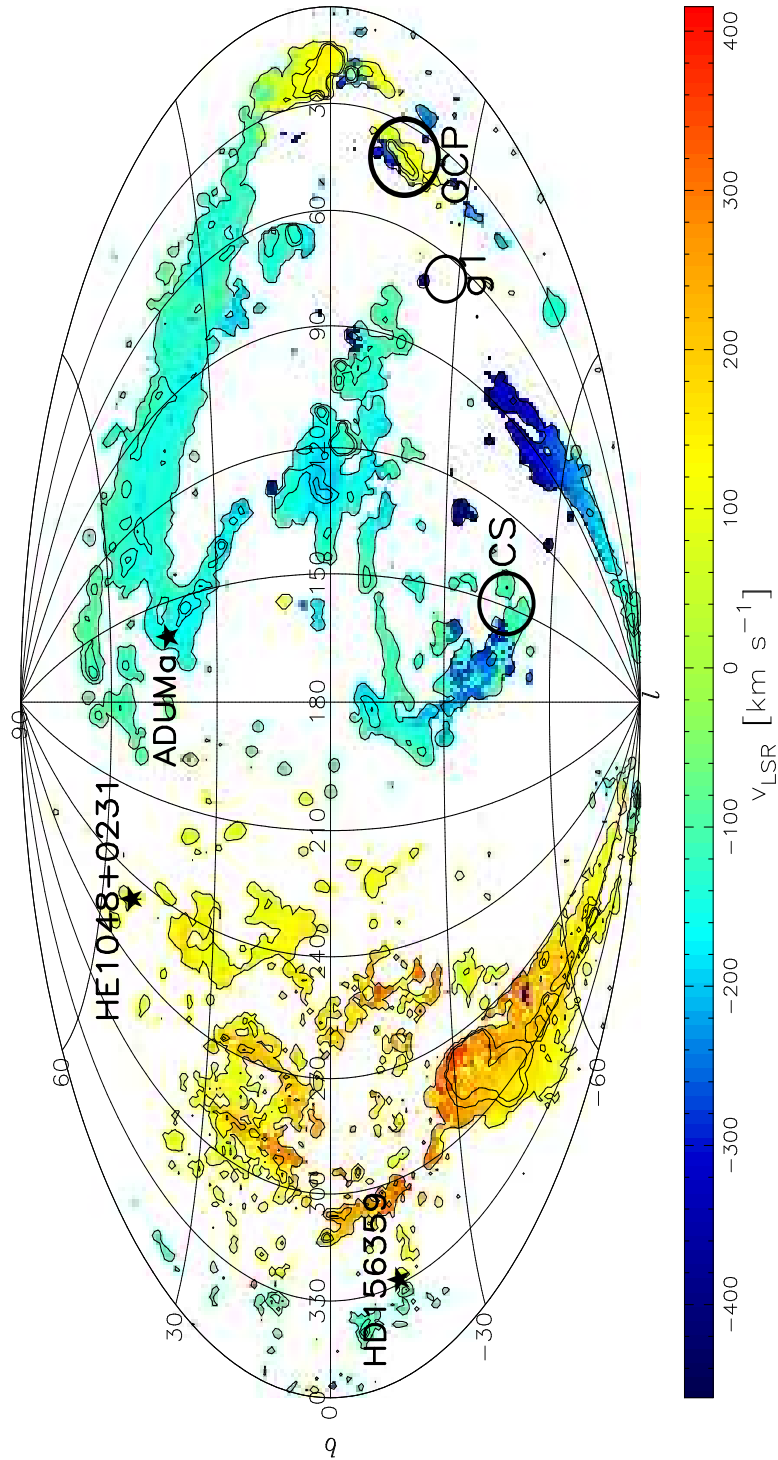}{0in}{0}{290}{550}{120}{0}\figurenum{1}
\caption{%
All-sky map of the high-velocity cloud sky, based on the data of Hulsbosch \&
Wakker (1988) and Morras et al.\ (2000). Colors represent LSR velocities, as
coded in the wedge. Contour levels are at brightness temperatures of 0.05, 0.25,
0.5 and 1~K. The positions of the clouds discussed in this paper are shown by
the thick-rimmed circles. Note that cloud~g1 does not show up in this map
because its velocity (+70~\kms) falls outside the velocity range covered by
Hulsbosch \& Wakker (1988). Also shown are the locations of the three stars
toward which high-velocity absorption was previously detected.
}\end{figure}

\begin{figure}\plotfiddle{f2.ps}{0in}{270}{350}{350}{30}{0}\figurenum{2}
\caption{%
Map of the column density of the Cohen Stream, integrating the LAB data of
Kalberla et al.\ (2005) between \vlsr=$-$130~\kms\ and \vlsr=$-$80~\kms.
Contours are at column densities of 1, 2 and 4\tdex{19}\,\cmm2. The locations of
the background stars are overlaid. The size (i.e.\ area) of the symbols is
proportional to the distance of the star. Closed circles are for directions
where the HVC is detected in absorption. Closed triangles are for significant
non-detections. Open circles are for non-detections that do not allow a
conclusion concerning the HVC's distance. Crosses are for stars where the
interstellar absorption is too blended with stellar lines. The QSO is shown by
an five-pointed star instead of a closed circle.
}\end{figure}

\begin{figure}\plotfiddle{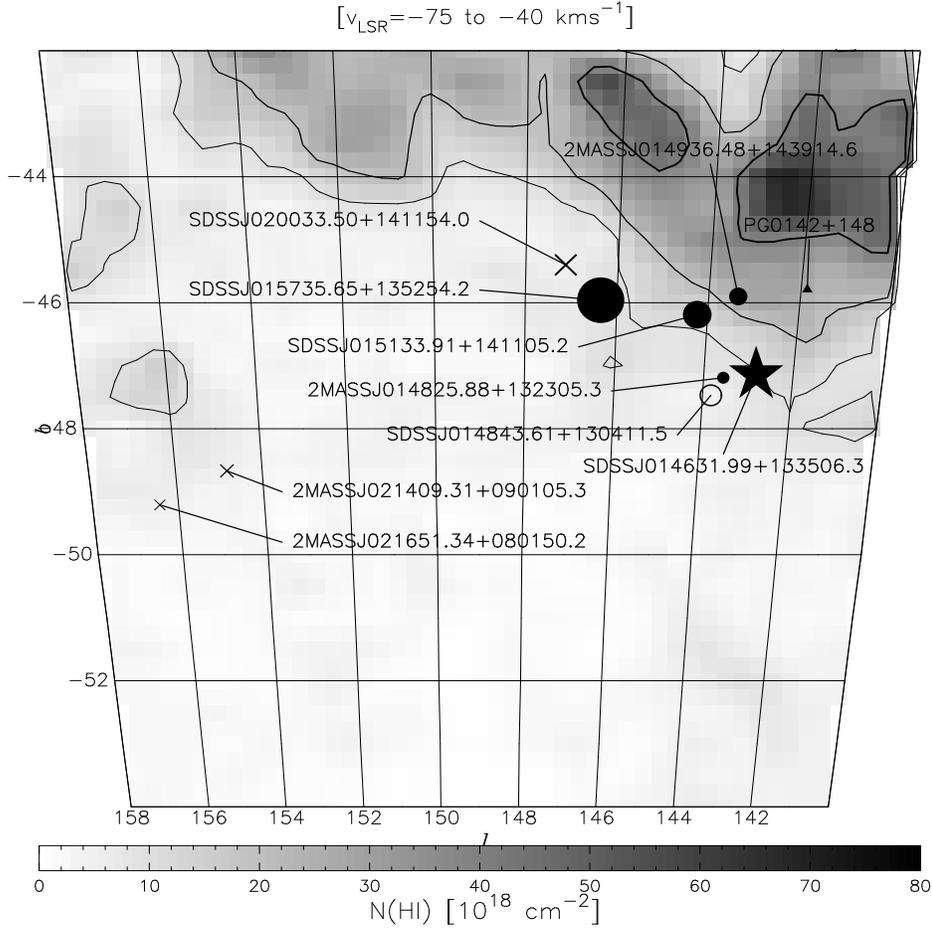}{0in}{270}{350}{350}{30}{0}\figurenum{3}
\caption{%
Map of the column density of the IVC in the Cohen Stream field, gas probably
associated with the high-z extension of the Perseus Arm, integrating the LAB
data of Kalberla et al.\ (2005) between \vlsr=$-$80~\kms\ and \vlsr=$-$30~\kms.
Contours are at column densities of 1, 2 and 4\tdex{19}\,\cmm2. The locations of
the background stars are overlaid. Symbols are as for Fig.~2.
}\end{figure}

\begin{figure}\plotfiddle{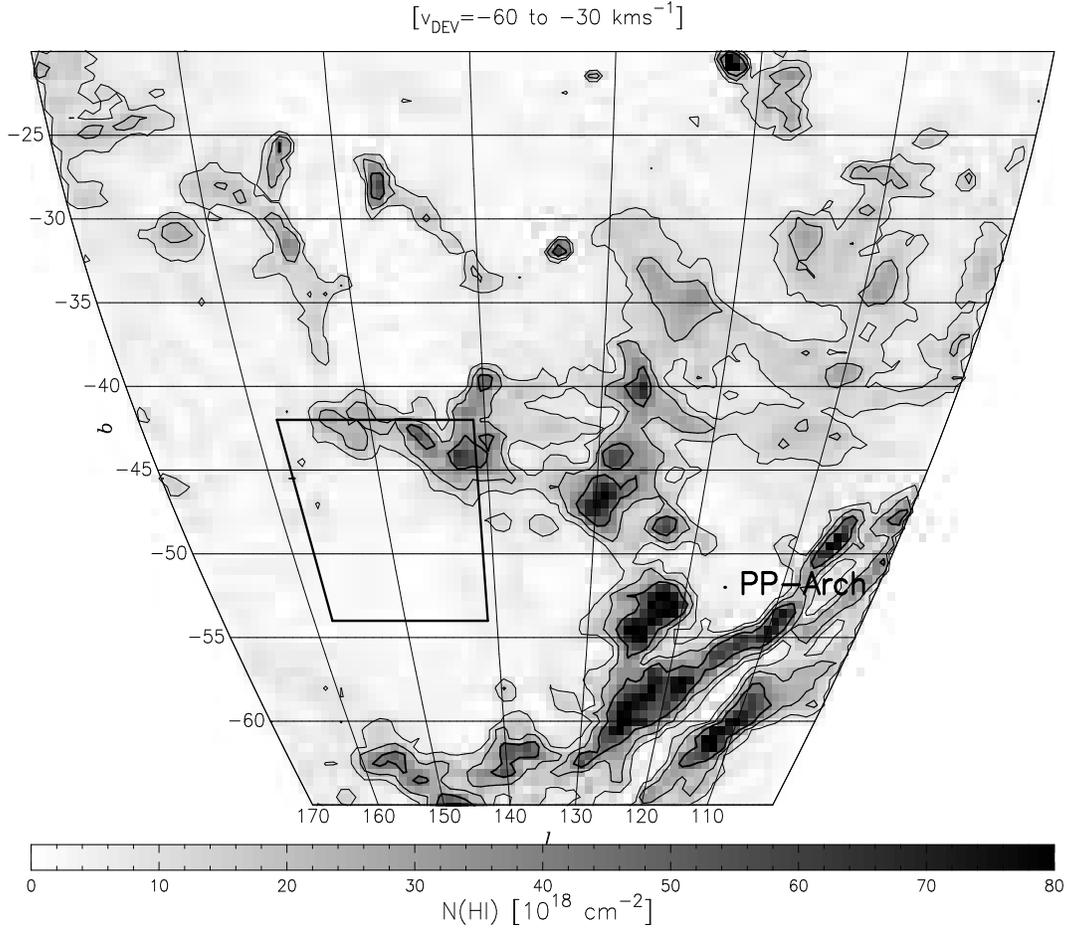}{0in}{270}{350}{400}{30}{-40}\figurenum{4}
\caption{%
Map of the column density of the intermediate-velocity gas in a large region to
the south of the Perseus Arm. This map represents the \HI\ column density in the
LAB data of Kalberla et al.\ (2005) integrated between deviation velocities
\vdev=$-$60~\kms\ and \vdev=$-$30~\kms. Contours are at column densities of 1, 2
and 4\tdex{19}\,\cmm2. The box outlines the region shown in Figs.~2 and 3.
}\end{figure}

\begin{figure}\plotfiddle{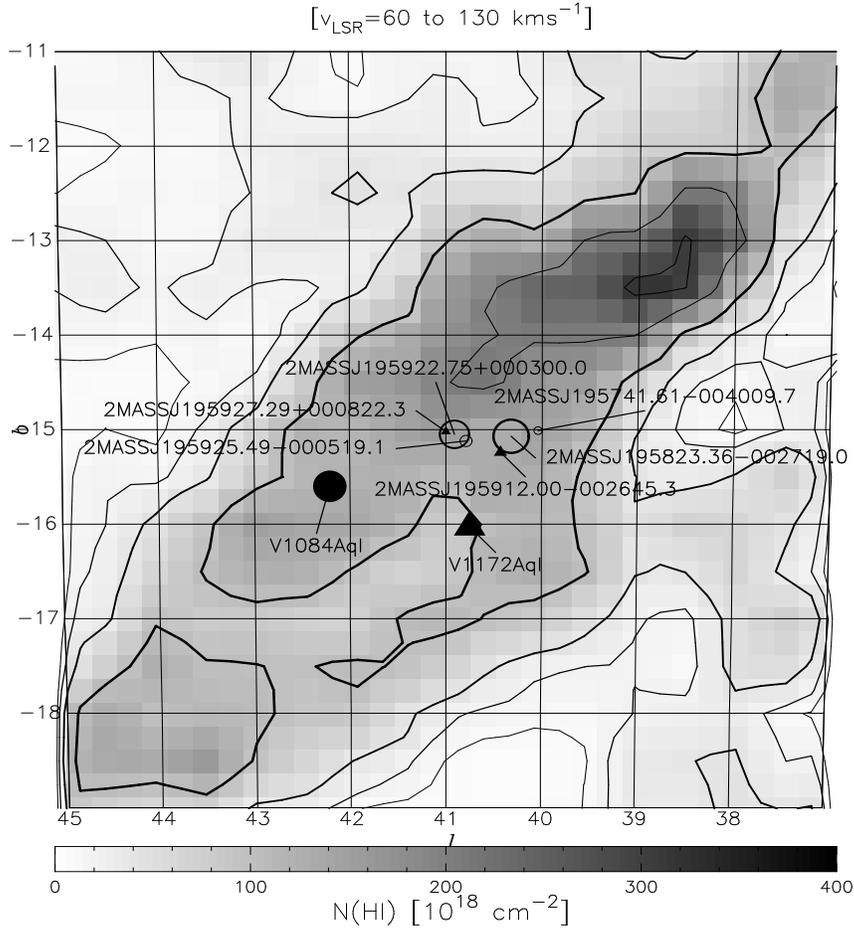}{0in}{270}{350}{320}{30}{0}\figurenum{5}
\caption{%
Map of the column density of complex~GCP, integrating the LAB data of Kalberla
et al.\ (2005) between \vlsr=+60~\kms\ and \vlsr=+130~\kms. Contours are at
$N$(\HI)= 1, 2, 4, 10, 20 and 30\tdex{19}\,\cmm2. Symbols are as for Fig.~2.
}\end{figure}

\begin{figure}\plotfiddle{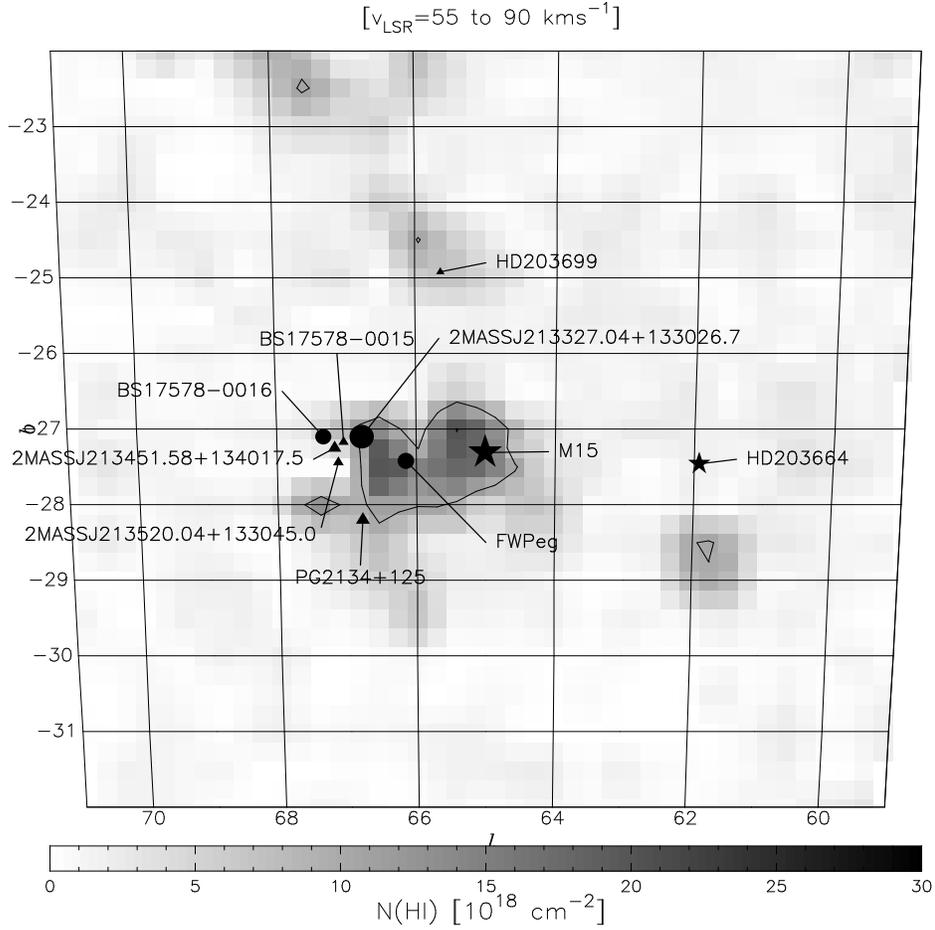}{0in}{270}{350}{350}{30}{0}\figurenum{6}
\caption{%
Map of the column density of the g1 cloud, integrating the LAB data of Kalberla
et al.\ (2005) between \vlsr=+55~\kms\ and \vlsr=+90~\kms. Contours are at 1, 2,
and 4\tdex{19}\,\cmm2. Symbols are as for Fig.~2.
}\end{figure}

\clearpage
\thispagestyle{empty}
\setlength{\voffset}{-15mm}
\begin{figure}\plotfiddle{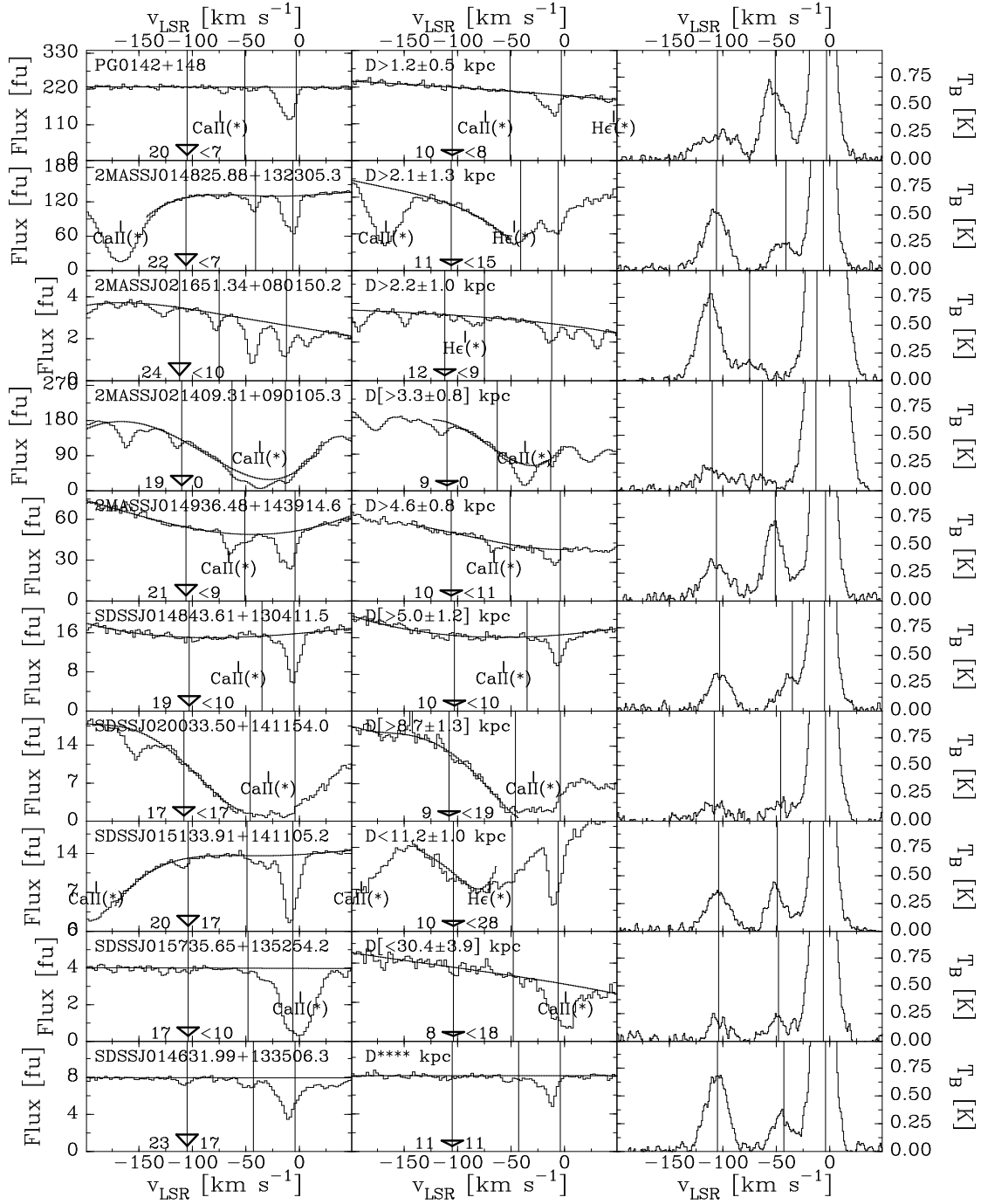}{0in}{0}{420}{520}{30}{20}\figurenum{7}
\caption{%
Spectra for stars toward the Cohen Stream. Flux units (f.u.) are \dex{-16} \fu.
Left column: \CaII~K spectra (histograms) and continuum fits (solid curves);
middle column: \CaII~H spectra and continua (note that the left flux scale is
valid only for \CaII~K); right column: \HI-21 cm spectra. One star per row, with
the name and distance given in the top left corner of the K and H panels.
Distances are preceded by ``$>$'' for lower limits, by ''$<$'' for upper limits.
Distances in brackets show that the detection limit is worse than the expected
equivalent width, or there are too many stellar lines. Labels (\CaII(*) and
\He(*)) show the positions of stellar absorption lines. Vertical lines give the
velocities of the \HI\ components. Triangles and numbers in each \CaII\ panel
give the shape of the expected absorption line, the expected equivalent width,
and the 3$\sigma$ detection limit or the detected equivalent width (a ``0''
means there is a blended stellar line).
}\end{figure}
\clearpage
\setlength{\voffset}{0mm}

\begin{figure}\plotfiddle{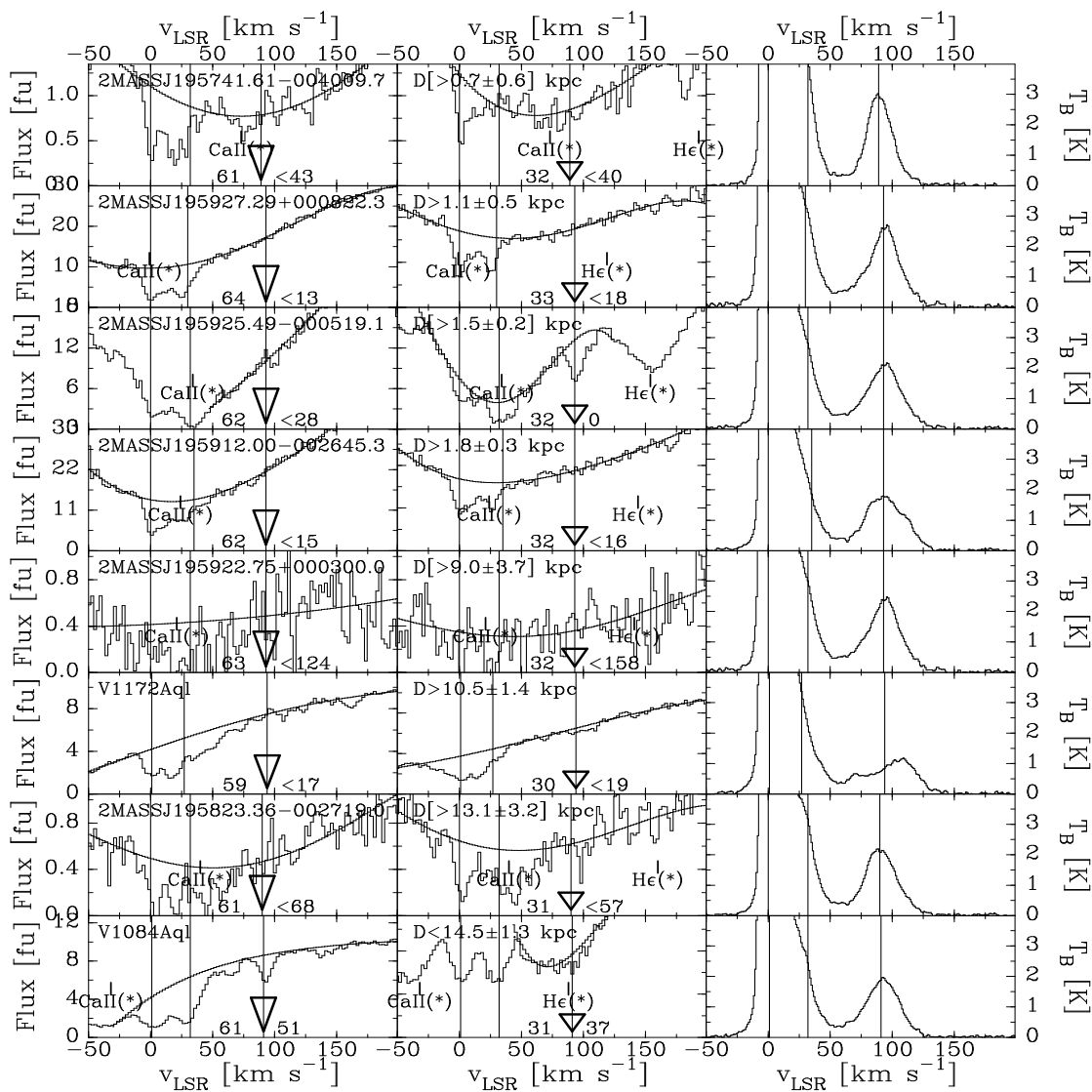}{0in}{0}{420}{420}{30}{20}\figurenum{8}
\caption{%
Same as Fig.~\FsCS, except for complex~GCP.
}\end{figure}

\begin{figure}\plotfiddle{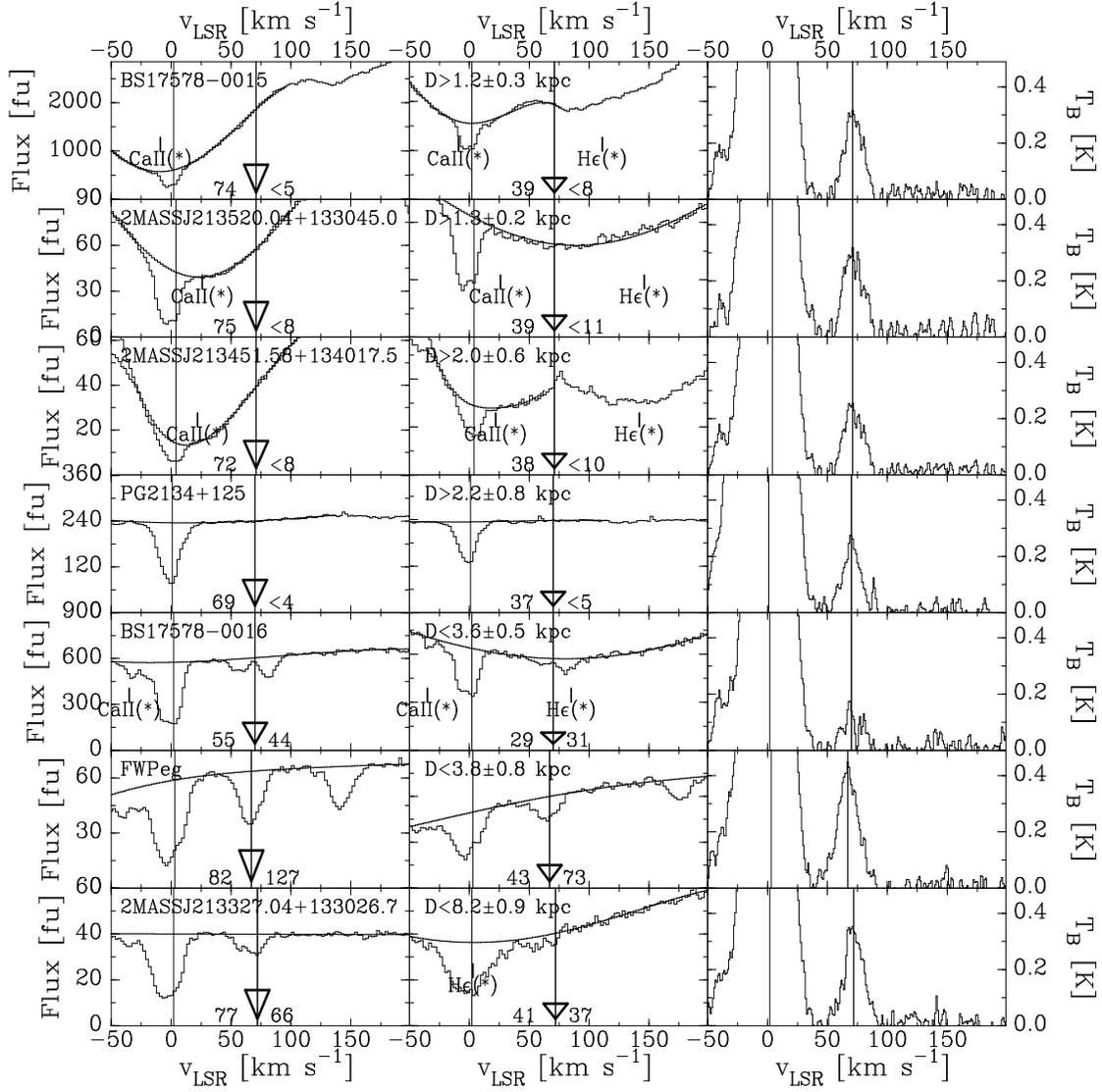}{0in}{0}{420}{420}{30}{20}\figurenum{9}
\caption{Same as Fig.~\FsCS, except for cloud~g1.
}\end{figure}

\end{document}